\documentclass[a4paper,11pt]{article}
\pdfoutput=1 

\usepackage{adjustbox}
\usepackage{amsfonts}
\usepackage{amssymb, amscd}
\usepackage{graphicx}
\usepackage{mathrsfs}
\usepackage[table,xcdraw]{xcolor}
\usepackage{colortbl}
\usepackage{slashed}
\usepackage{pdfpages}
\usepackage{braket}
\usepackage{sidecap}
\usepackage{arydshln}
\usepackage[font=small,labelsep=none]{caption}
\usepackage{mathtools}
\usepackage{hyperref}
\usepackage{tikz}
\usetikzlibrary{arrows,decorations.markings}
\usepackage{subcaption}
\usepackage[left=2cm,right=2cm,top=3.5cm,bottom=3.5cm]{geometry}

\hypersetup{
    colorlinks=true,
    linkcolor=ceruleanblue,
    filecolor=ceruleanblue,      
    urlcolor=ceruleanblue,
    citecolor=ceruleanblue,
}

\definecolor{myblue}{RGB}{174, 198, 219}
\definecolor{myred}{RGB}{157,31,68}
\definecolor{ceruleanblue}{rgb}{0.0, 0.2, 0.6}

\newcommand{\be}{\begin{equation}}      
\newcommand{\ee}{\end{equation}}

\numberwithin{equation}{section}

\def\MP{M_{\rm Pl}}
\newcommand*{\pt}{\partial}%
\newcommand{\vect}[1]{\boldsymbol{#1}}
\newcommand{\phih}{h}
\newcommand{\pih}{\phi}
\newcommand{\varphih}{\delta h}

\newcommand{\eqn}[1]{eq.~(\ref{#1})}

\begin{document}


\begin{center}
\LARGE{\bf Dark-Energy Instabilities induced by\\ Gravitational Waves}
\\[1cm] 

\large{Paolo Creminelli$^{\,\rm a, \rm b}$, Giovanni Tambalo$^{\,{\rm c},{\rm d}}$, Filippo Vernizzi$^{\, \rm e}$ and Vicharit Yingcharoenrat$^{{\,\rm c }, {\rm d}}$}
\\[0.5cm]

\small{
\textit{$^{\rm a}$
ICTP, International Centre for Theoretical Physics\\ Strada Costiera 11, 34151, Trieste, Italy}}
\vspace{.2cm}

\small{
\textit{$^{\rm b}$
IFPU - Institute for Fundamental Physics of the Universe,\\ Via Beirut 2, 34014, Trieste, Italy }}
\vspace{.2cm}

\small{
\textit{$^{\rm c}$ SISSA, via Bonomea 265, 34136, Trieste, Italy}}
\vspace{.2cm}

\small{
\textit{$^{\rm d}$ INFN, National Institute for Nuclear Physics \\  Via Valerio 2, 34127 Trieste, Italy}}
\vspace{.2cm}

\small{
\textit{$^{\rm e}$ Institut de physique th\' eorique, Universit\'e  Paris Saclay, CEA, CNRS \\ [0.05cm]
91191 Gif-sur-Yvette, France}}
\vspace{.2cm}

\end{center}

\vspace{0.3cm}


\vspace{0.3cm}

\begin{abstract}\normalsize
We point out that dark-energy perturbations may become unstable in the presence of a gravitational wave of sufficiently large amplitude. We study this effect for the cubic Horndeski operator (braiding), proportional to $\alpha_{\rm B}$. The scalar that describes dark-energy fluctuations features ghost and/or gradient instabilities for gravitational-wave amplitudes that are produced by typical binary systems. Taking into account the populations of binary systems, we conclude that the instability is triggered in the whole Universe for $|\alpha_{\rm B} |\gtrsim 10^{-2}$, i.e.~when the modification of gravity is sizeable. The instability is triggered by massive black-hole binaries down to frequencies corresponding to $10^{10}$ km: the instability is thus robust, unless new physics enters on even longer wavelengths. The fate of the instability and the subsequent time-evolution of the system depend on the UV completion, so that the theory may end up in a state very different from the original one. The same kind of instability is present in beyond-Horndeski theories for $|\alpha_{\rm H}| \gtrsim 10^{-20}$. 
In conclusion, the only dark-energy theories with sizeable cosmological effects that avoid these problems are $k$-essence models, with a possible conformal coupling with matter.  

\end{abstract}

\vspace{0.3cm} 


\newpage 
{
 \hypersetup{linkcolor=black}
 \tableofcontents
}

\vspace{0.3cm}

\flushbottom



\section{Introduction}\label{sec:intro}

In modified gravity, gravitational waves (GWs) can decay into scalar field fluctuations  \cite{Creminelli:2018xsv,Creminelli:2019nok}, inducing an observational signature in ground- and space-based interferometers such as LIGO-Virgo \cite{LIGOColl} and LISA \cite{Audley:2017drz}.  
This only happens in models where Lorentz invariance is  broken spontaneously, such as for instance in scalar-tensor gravity  with a homogeneous cosmological scalar field, motivated by the accelerated expansion of the Universe. 
For the effect to be sizeable, one needs a  cubic coupling $\gamma \pi \pi$ ($\gamma$ denotes the GW and $\pi$ a scalar field fluctuation) suppressed by a  sufficiently low energy scale. 
For instance, this coupling is present in theories beyond Horndeski \cite{Zumalacarregui:2013pma,Gleyzes:2014dya}. The scale $\Lambda_\star$ that suppresses the higher-dimension operators is of order $\Lambda_3 \equiv (H_0^2 \MP)^{1/3}$, in the regime where one gets sizeable effects for the formation of structures in the Universe.

In \cite{Creminelli:2018xsv} we studied the decay induced by this interaction {\em perturbatively}, i.e.~when individual gravitons decay independently of each other, and we showed that the absence of perturbative decay implies that $\Lambda_\star \gtrsim 10^3 \Lambda_3$,  setting a tight bound on the parameter space of these models. In particular, in the Effective Field Theory of Dark Energy (EFT of DE)  \cite{Creminelli:2006xe,Cheung:2007st,Creminelli:2008wc,Gubitosi:2012hu,Gleyzes:2013ooa} formalism  and focussing on theories with GWs that propagate luminally,  beyond-Horndeski theories  are characterised by a single operator: $\frac12 \tilde m_4^2(t) \delta g^{00} \left({}^{(3)}\!R + \delta K_\mu^\nu \delta K_\nu^\mu - \delta K^2  \right) $. (Details on the quantities appearing in this formula will be given below, in Secs.~\ref{sec:cubic_action} and \ref{sec:m4}.) The absence of decay sets a
bound on this operator: $|\tilde m_4^2| \lesssim 10^{-10} \MP^2 $. Equivalently, in terms of the dimensionless parameter introduced in \cite{Gleyzes:2014qga},  this translates in the bound $| \alpha_{\rm H}| \lesssim 10^{-10}$.\footnote{More specifically, this constraint applies only for GLPV theories. For more general theories beyond Horndeski, such as the Degenerate-Higher-Order-Scalar-Tensor (DHOST) theories \cite{Langlois:2015cwa,Crisostomi:2016czh}, the constraint becomes $\alpha_{\rm H} + 2 \beta_1 \lesssim 10^{-10}$ \cite{Creminelli:2018xsv}, where $\beta_1$ characterizes higher-order operators in the EFT of DE parameterization \cite{Langlois:2017mxy}. The consequences of this constraint on the Vainshtein mechanism in these theories has been studied in \cite{Hirano:2019scf,Crisostomi:2019yfo}.} This rules out the possibility of observing the effects of these theories in the large-scale structure.
 
In \cite{Creminelli:2019nok} we extended this study to consider coherent effects due to the large occupation number of the GW, acting as a classical background for $\pi$. 
In this case, a better description of the system is that of parametric resonance: $\pi$ fluctuations  
are described by a Mathieu equation and are exponentially produced by parametric instability.
We focused on the regime of {\em narrow resonance}, obtained when the GW induces a small perturbation on the $\pi$ equation.\footnote{The calculation in the narrow resonance regime reduces to the perturbative one when the occupation number is
small enough \cite{Creminelli:2019nok}.} This regime can be used to probe only very small values of $\alpha_{\rm H}$. In particular,  within the validity of our approximations the resonant decay takes place in the range $10^{-20} \lesssim | \alpha_{\rm H}| \lesssim 10^{-17}$ for frequencies of interest for LIGO-Virgo and  $10^{-16} \lesssim| \alpha_{\rm H}| \lesssim 10^{-10}$ for LISA.  

Another operator containing the cubic coupling $\gamma \pi \pi$ is $m_3^3 (t) \delta g^{00} \delta K$ \cite{Creminelli:2006xe,Creminelli:2008wc}. In the covariant language, it corresponds to the cubic Horndeski Lagrangian  (sometimes called kinetic gravity braiding \cite{Deffayet:2010qz,Kobayashi:2010cm}). In this case, the scale $\Lambda$ that suppresses this cubic interaction is typically much higher than $\Lambda_3$, i.e.~$\Lambda \sim \Lambda_2 \equiv (H_0 \MP)^{1/2}$ and the perturbative decay is negligible \cite{Creminelli:2018xsv}. Moreover, non-linearities in the dark energy field become sizeable much before the effect of narrow resonance  is relevant, possibly quenching the coherent   instability \cite{Creminelli:2019nok}. Therefore, the study of the perturbative and resonant decay for this operator remains inconclusive. 

In this article we study the effect of a classical GW on  $\pi$  in the regime where the amplitude of the  wave is large, i.e.~far from the narrow resonance, focussing on the stability of $\pi$ perturbations.
We initially concentrate on the operator $m_3^3$, introducing the action and setting up  the notation in Sec.~\ref{sec:cubic_action}. (Deviations from this case are studied in App.~\ref{app:cG}.)
Inspired by the analysis of Ref.~\cite{Nicolis:2004qq} reviewed in Sec.~\ref{sec:NicolisRattazzi}, in the rest of Sec.~\ref{sec:stabArg} we 
compute the non-linear classical solution of $\pi$ generated by the GW and we study the stability of $\pi$ fluctuations, outlining the differences with the analysis of \cite{Nicolis:2004qq}. We consider  two different regimes: subluminal  and luminal
speed of $\pi$ fluctuations,  respectively examined in Secs.~\ref{sec:csless1} and \ref{sec:csone}. Both cases display instabilities and qualitatively agree.

Without the knowledge of the UV completion of the theory, we cannot describe the evolution of the system and its endpoint. We discuss this issue in Sec.~\ref{sec:fateofin} with an example that displays similar instabilities and whose UV completion is known. Anyway, the theory must change qualitatively in the regions where the instability develops.
In Sec.~\ref{sec:constraints} we study whether the populations of binary systems and their production of GWs is enough to trigger the instability in the whole Universe. Stellar and massive black holes (BHs) are able to globally induce the instability in the regime where one has a sizeable effect on structure formation ($|\alpha_{\rm B}| \gtrsim 10^{-2}$). The instability is triggered by GWs as long as $10^{10}$ km, so that our conclusions are robust unless the theory is modified on even longer scales.
In Sec.~\ref{sec:m4}, we discuss the application of our study to the operator $\tilde m_4^2$ as well, and we derive strong bounds of order $|\alpha_{\rm H}|\lesssim 10^{-20}$.
Finally, we discuss our conclusions and future prospects in Sec.~\ref{sec:con}.

\section{The action}\label{sec:cubic_action}

We consider the following action in unitary gauge, 
\begin{equation}
\label{starting_action}
S =  \int  \text d^4 x \sqrt{-g}  \left[  \frac{\MP^2}{2}  \, {}^{(4)}\!R - \lambda(t)- c(t) g^{00}  +  \frac{m_2^4(t)}{2} (\delta g^{00})^2  - \frac{m_3^3(t)}{2}\delta g^{00} \delta K -\frac{\tilde m_3^3(t)}{8}(\delta g^{00})^2 \delta K \right]\,,
\end{equation} 
and focus in particular on the  cubic Galileon, i.e.~$\tilde m_3^3 = - m_3^3$  \cite{Gubitosi:2012hu}. Generalizations of this case are discussed in App.~\ref{app:cG}. We took a constant $\MP$ with a proper choice of frame and we will discuss the general case in the conclusions. Here we assumed the flat FRW metric, $\text d s^2 = - \text d t^2 + a^2(t) \text d \vect x^2 $, so that $\delta g^{00}\equiv 1+g^{00}$ is the perturbation of $g^{00}$ around the background solution. We also defined the perturbation of the extrinsic curvature of the equal-time hypersurfaces as $\delta K^\mu_{\ \nu} \equiv K^\mu_{\ \nu}-H\delta^\mu_{\ \nu}$, where $H \equiv \dot a/a$ is the Hubble rate, and its trace as $\delta K$.
As discussed below, the operator proportional to $m_2^4$ affects the quadratic action for $\pi$, contributing to the overall normalization of the action and to the speed of sound.   
It also introduces self-interactions but in the cosmological setting these are suppressed by $\Lambda_2 \gg \Lambda_3$ and can be dropped for this discussion, because they are irrelevant for the stability. For the same reason, we can ignore higher powers of $\delta g^{00}$.

We restore the $\pi$ dependence in a generic gauge with the Stueckelberg trick \cite{Cheung:2007st,Gleyzes:2013ooa} $t \rightarrow t+\pi(t,\vect{x})$. We focus on terms relevant for our calculations, i.e.~\cite{Cusin:2017mzw}
 \begin{align}
 g^{00} &\rightarrow g^{00}+2g^{0\mu}\partial_\mu\pi+g^{\mu\nu}\partial_\mu\pi\partial_\nu\pi \;,\label{eq:Stuekelberg1}\\ 
 \delta K &\rightarrow \delta K -h^{ij}\partial_i\partial_j\pi + \frac{2}{a^2} \partial_i \pi \partial_i \dot \pi +\ldots \; . \label{eq:Stuekelberg2} 
 \end{align}

We follow \cite{Creminelli:2018xsv,Creminelli:2019nok} and work in Newtonian gauge, where the line element reads
 \begin{equation}\label{metric:Newtonian}
 \textrm ds^2 = -(1+2\Phi)\textrm dt^2 + a^2(t)(1-2\Psi)(e^\gamma)_{ij}\textrm dx^i \textrm dx^j \ ,
 \end{equation}
 with $\gamma$ transverse, $\partial_i \gamma_{ij} =0$, and traceless, $\gamma_{ii} =0$.
 Variation of the action with respect to $\Phi$ and $\Psi$ gives, focussing on the sub-Hubble limit by keeping only the leading terms in spatial derivatives, \cite{Creminelli:2018xsv,Creminelli:2019nok} 
\begin{align}\label{eq:constraints}
    \Phi = \Psi = -\frac{m_3^3}{2 \MP^2}\pi \;.
\end{align}
These relations can be used to replace $\Phi$ and $\Psi$ in terms of $\pi$ in the action.

As in \cite{Creminelli:2019nok}, we define the dimensionless quantity \cite{Bellini:2014fua,Gleyzes:2014rba}
\be\label{alphaB}
\alpha \equiv \frac{4 \MP^2 (c+ 2 m_2^4)+ 3 m_3^6}{2 \MP^4 H^2}   
 \;, 
  \ee
and we canonically normalise scalar and tensor perturbations as
 \begin{equation}\label{canonical}
 \pi_c \equiv  \sqrt{\alpha} \MP H \pi \;,\qquad \gamma_{ij}^{c} \equiv \frac{\MP}{\sqrt{2}}\gamma_{ij} \;.
 \end{equation}
In the following we will use canonically normalized  fields but we will drop the symbol of canonical normalizations. Neglecting the expansion, the action for $\pi$ reads
$ S_\pi = \int \text d^4x {\mathscr L } $,
where the $\pi$ Lagrangian is
\be
  \label{InteractionLagrangian}
 \mathscr L = 
 \frac{1}{2} \left[ \dot{\pi}^2 - {c_s^2}(\partial_i\pi)^2 \right] -\frac{1}{\Lambda_{\rm B}^3} \square \pi (\partial \pi)^2 + \frac{1}{\Lambda^2}\dot{\gamma}_{ij}\partial_i\pi\partial_j\pi  + \frac{m_3^3}{2 \sqrt{\alpha} \MP^3 H}\pi\dot{\gamma}_{ij}^2 \; ,
\end{equation}
where $\square \pi \equiv \eta^{\mu\nu}\partial_\mu\partial_\nu \pi$ and $(\partial \pi)^2 \equiv \eta^{\mu \nu}\partial_\mu\pi \partial_\nu\pi$.
Here we used the definitions
\begin{align}
 c_s^2 &\equiv \frac{4\MP^2c-m_3^3(m_3^3-2\MP^2H)}{4\MP^2(c+2m_2^4) + 3m_3^6}   = \frac{2}{\alpha} \left( \frac{c}{\MP^2H^2} - \alpha_{\rm B}- \alpha_{\rm B}^2 \right) \;, \label{Cs} \\
 \Lambda^2 & \equiv \frac{4\MP^2(c+2m_2^4) + 3m_3^6}{\sqrt{2}m_3^3\MP}  = - \frac{\alpha}{\sqrt{2} \alpha_{\rm B}} \frac{H}{H_0} \Lambda_2^2\ \label{InterScale} \;, \\
\Lambda_{\rm B}^3 & \equiv - \frac{[ 3 m_3^6 + 4 \MP^2 (c+ 2 m_2^4)]^{3/2} }{\sqrt{2} m_3^3  \MP^3 }  = \frac{{\alpha}^{3/2}}{\alpha_{\rm B}} \left( \frac{H}{H_0} \right)^2 \Lambda_3^3\;.  \label{Mtre}
\end{align}
On the right-hand side of these relations, we have given the analogous definitions in terms of $\Lambda_2 \equiv (H_0 \MP)^{1/2}$, $\Lambda_3 \equiv (H_0^2 \MP)^{1/3}$ and the dimensionless quantity $\alpha_{\rm B}$, here defined as
\be\label{alphaB}
 \alpha_{\rm B} \equiv  -\frac{m_3^3}{2 \MP^2 H} \;.
\ee
 
The detailed calculation of the above Lagrangian can be found  in \cite{Creminelli:2018xsv,Creminelli:2019nok}, except for the last operator,  
containing  the cubic coupling $\gamma \gamma \pi$.
 Such vertex is not directly obtained  from the operator $m_3^3 \delta g^{00} \delta K$. Instead, it comes from the Einstein-Hilbert term of the action \eqref{starting_action} when replacing the potentials $\Phi$ and $\Psi$ with $\pi$ {\em via}  \eqn{eq:constraints}.\footnote{Since we know that a tensor perturbation $\gamma_{ij}$ couples with the metric in the same way as a scalar field does, one can easily obtain this interaction by considering the Lagrangian of minimally coupled scalar field and  replacing the scalar field by $\gamma_{ij}$.}

For the GW, we will use the classical background solution travelling in the $\hat z$ direction with linear polarization $+$,  used in \cite{Creminelli:2019nok},
\be \label{gamma}
\gamma_{ij} = \MP h_0^+ \sin \left[ \omega (t-z) \right] \epsilon_{ij}^+ \;,
\ee
where $h_0^+$ is the dimensionless strain amplitude and $\epsilon_{ij}^+ = \text{diag}(1,-1,0)$.
For later convenience, we also define the parameter \cite{Creminelli:2019nok}
\be \label{beta}
\beta \equiv \frac{2 \omega \MP h_0^+}{c_s^2 |\Lambda^2|} = \frac{2 \sqrt{2} |\alpha_{\rm B}|}{\alpha c_s^2} \frac{\omega}{H} h_0^+\;.
\ee

\section{Classical solutions and stability of perturbations}\label{sec:stabArg}
In any non-linear theory one can investigate the stability of a given solution by looking at the kinetic term of small perturbations around it. This was done for the cubic Galileon (equivalent to the decoupling limit of the Dvali-Gabadadze-Porrati (DGP) model) in \cite{Nicolis:2004qq}, in the absence of GWs. It was proven that solutions that are stable at spatial infinity are stable everywhere, provided the sources are non-relativistic. The analysis was later extended to higher Galileons in \cite{Endlich:2011vg}, where such strong statement does not hold and one expects that general non-linear solutions feature instabilities. Here we want to extend the analysis of \cite{Nicolis:2004qq} including GWs\footnote{As discussed in \cite{Bose:2018orj}, the DGP model  is not a local theory of a scalar field and thus is not included in the ordinary EFT  of DE action. However, the structure of the non-linear terms is analogous and the arguments used in \cite{Nicolis:2004qq} can be applied straightforwardly to the EFT of DE. On the other hand the brane-bending mode in the DGP model is not a scalar under 4d diffs \cite{Luty:2003vm}, so that the coupling with GWs will be different from the one discussed in this paper. } and considering a generic speed of propagation $c_s$. (In order to compare with the result of \cite{Nicolis:2004qq}, in the main text we stick to the non-linearity of the cubic Galileon, i.e.~$\tilde m_3^3 = - m_3^3$. In Appendix \ref{app:cG} we generalise the analysis to the case $\tilde m_3^3 \neq - m_3^3$.)

For convenience we define $\bar \eta_{\mu \nu} \equiv \text{diag} (-1, c_s^2, c_s^2, c_s^2)$ and  $\bar \square \pi \equiv \bar \eta^{\mu \nu}\partial_\mu \partial_\nu \pi = - \ddot{ \pi} + c_s^2 \partial_k^2  \pi$. In this section, indices are raised and lowered with the usual Minkowski metric.
Moreover, we define
\be
\Gamma_{\mu \nu} \equiv  \frac{\dot \gamma_{\mu \nu} }{ \Lambda^{2}} \;.
\ee
Using the above definitions, the action for  $\pi$, eq.~\eqref{InteractionLagrangian}, becomes
\begin{equation}\label{eq:LagDGPDet1}
\mathscr L = -\frac{1}{2}\bar \eta^{\mu \nu} \partial_\mu \pi \partial_\nu  \pi -\frac{1}{\Lambda_{\rm B}^3} \square \pi (\partial \pi)^2 + \Gamma_{\mu \nu} \partial^\mu \pi  \partial^\nu \pi  -  \frac{1}{2} \Lambda_{\rm B}^3 \pi \Gamma_{\mu \nu}^2 \;.
\end{equation}
In the following we will use that  $\partial^\mu \Gamma_{\mu \nu} = \eta^{\mu \nu} \Gamma _{\mu \nu}  = \bar \eta^{\mu \nu} \Gamma _{\mu \nu} = 0$.
For our GW solution \eqref{gamma} we have
\be
\label{Gamma}
\Gamma_{00} = \Gamma_{0i} =0 \;, \qquad \Gamma_{ij} = \frac{\beta c_s^2}{2} \cos \left[ \omega (t - z) \right] \epsilon_{ij}^+ \;,
\ee
where we have used the definition of $\beta$,  eq.~\eqref{beta}. The third term of eq.~\eqref{eq:LagDGPDet1} suggests that for $\beta > 1$ the scalar $\pi$ features a gradient instability, because $\Gamma_{\mu\nu}$ changes sign in time. This conclusion, although substantially correct, is premature, since the GW also sources a background for $\pi$ and this affects through non-linearities the behaviour of perturbations.

Let us split the field in a classical background part plus fluctuations, i.e.~$\pi = \hat \pi (t,\vect x) + \delta \pi (t, \vect x)$.
In general $\hat \pi$ will be sourced by the term $\dot \gamma_{\mu \nu}^2$ of \eqn{InteractionLagrangian}, corresponding to the last term of eq.~\eqref{eq:LagDGPDet1}, and also by astrophysical matter sources. 
Let us first study the background solution $\hat \pi$. Its equation of motion reads
\begin{equation}\label{eq:stab1}
\bar \square \hat  \pi - \frac{2}{\Lambda_{\rm B}^3} \left[ (\partial_\mu \partial_\nu\hat \pi )^2 - \square \hat \pi^2 \right] - 2 \Gamma_{\mu \nu}  \partial^\mu \partial^\nu \hat \pi   - \frac{\Lambda_{\rm B}^3}{2} \Gamma_{\mu \nu}^2 = 0 \;.
\end{equation}
Following \cite{Nicolis:2004qq}, we define
the matrix 
\be
{\cal K}_{\mu \nu} \equiv -  \frac{1}{\Lambda_{\rm B}^3} \partial_\mu \partial_\nu \hat \pi  \;,
\ee
and rewrite the above equation  as
\begin{equation}\label{eq:K_eq1}
{\cal K}^{\mu \nu}\bar \eta_{\mu \nu} + 2 \left( {\cal K}_{\mu \nu} {\cal K}^{\mu \nu} - {\cal K}^2 \right) - 2 \Gamma_{\mu \nu} {\cal K}^{\mu \nu}  + \frac{1}{2}\Gamma_{\mu \nu}^2= 0 \;.
\end{equation}
Due to the Galileon symmetry and the fact the equations of motion are second order, eq.~\eqref{eq:stab1} reduces to 
an algebraic equation for the second derivatives  of $\hat \pi$. Eq.~\eqref{eq:K_eq1} can be rewritten solely in terms of $\mathcal{\tilde{K}_{\mu\nu}} \equiv \mathcal{K}_{\mu\nu} - \frac{1}{2}\Gamma_{\mu\nu}$, and becomes
\begin{equation}\label{eq:K_eq2}
{\cal \tilde K}^{\mu \nu}\bar \eta_{\mu \nu} + 2 \left( {\cal \tilde K}_{\mu \nu} {\cal \tilde K}^{\mu \nu} - {\cal \tilde K}^2 \right) = 0 \;.
\end{equation}

The stability of a generic solution of eq.~\eqref{eq:K_eq1} can be assessed by studying the quadratic Lagrangian for the perturbations $\delta \pi$. These are assumed to be of a wavelength much shorter than the typical variation of $\hat \pi$.
Expanding the action \eqref{eq:LagDGPDet1} at quadratic order in $\delta \pi$, after some integrations by parts we obtain
\begin{equation}\label{eq:lagpert}
\mathscr L_{(2)} =  Z^{\mu \nu}(x) \, \partial_\mu \delta \pi \partial_\nu  \delta \pi \;,  \qquad Z^{\mu \nu} \equiv - \frac{1}{2}\bar \eta^{\mu \nu} - 2 \left({\cal \tilde K}^{\mu \nu} - \eta^{\mu \nu} {\cal \tilde K} \right) \;,
\end{equation}
where the indexes are raised and lowered with the Minkowski metric $\eta_{\mu \nu}$.
In general, for time-dependent kinetic terms like \eqref{eq:lagpert} there is no clear definition of stability. However, in the limit that we consider here where $Z^{\mu \nu}(x)$ changes much slower  
than the fluctuations $\delta \pi$, the requirement of stability simply translates in the absence of ghost or gradient instabilities for the perturbations, i.e.~$Z^{00} >0$ and that $Z^{0i}Z^{0j} - Z^{ij}Z^{00}$ is a positive-definite matrix at each point \cite{Nicolis:2004qq,Dubovsky:2005xd}. As explained in \cite{Dubovsky:2005xd}, a theory can be stable even when $Z^{00}<0 $, provided it features superluminal excitations and one can boost to a frame in which $Z^{00} >0$.\footnote{In the absence of superluminality the sign of $Z^{00}$ is invariant under Lorentz transformations. The positive definiteness of the matrix $Z^{0i}Z^{0j} - Z^{ij}Z^{00}$ is always invariant.} However, in our problem we have a privileged frame, the cosmological one, where the Cauchy problem must be well defined. Therefore stability must be manifest in this specific frame.

It is important to stress that the Newtonian gauge is very special for our analysis. In this paper we are interested in a fully non-linear analysis, but we solved for $\Psi$ and $\Phi$ linearly, see eq.~\eqref{eq:constraints}. This is justified in Newtonian gauge because, even when the equation of motion of $\pi$ becomes non-linear, $\Phi$ and $\Psi$ remain small and higher-order terms can be neglected. (This is analogous to what happens for non-linearities in the Large-Scale Structure: perturbation theory for the density contrast $\delta$ breaks down on short scales, but $\Phi$ and $\Psi$ remain small and perturbative.) This does not happen in other gauges. For instance, in spatially-flat gauge one can take the solution for the shift function, see eq.~\eqref{eq:constraint_psi} in App.~\ref{app:B}. In the regime of interest $\alpha_\psi \sim 1$ and Galileon non-linearities are relevant for $\partial^2\pi_c \sim H^2 \MP$, which implies $\partial^2\pi \sim H$. Therefore, the perturbation in the extrinsic curvature is $\delta K \sim \nabla^2 \psi \sim H$, which is of the same order as the background value. Thus higher-order corrections in the constraint equations become relevant, since the Einstein-Hilbert action contains terms quadratic in the extrinsic curvature that cannot be neglected. A similar behaviour occurs in comoving gauge. The analysis in these gauges is therefore much more complicated. As a partial check of our calculation, in App.~\ref{app:B} we verify that our Newtonian action matches the action in spatially flat gauge, but we do this only at the perturbative level, at cubic order. 

It is important to stress that, although our analysis is done in a particular gauge, the matrix ${\cal \tilde K}_{\mu \nu}$ is a covariant tensor: 
\be
\mathcal{\tilde{K}_{\mu\nu}} = \mathcal{K_{\mu\nu}} - \frac{1}{2}\Gamma_{\mu\nu} =-\frac{\alpha_{\rm B}}{\alpha H} \cdot \nabla_\mu \nabla_\nu \phi \;,
\ee
where $\phi \equiv t + \pi$ is the complete dark-energy scalar field (not in canonical normalization) and it is a scalar quantity under all diffs. (The second equality works only if we neglect non-linear terms involving $\pi$ and Christoffel's symbols: one can check that these are subdominant with respect to the terms we kept.) Therefore the matrix $Z^{\mu\nu}$ is a covariant tensor\footnote{Actually, $\bar\eta^{\mu\nu}$ depends on $\pi$ perturbations but again the terms that we are neglecting are subdominant with respect to the ones we kept.} and the conditions for stability are gauge independent. 

The matrix $Z_{\mu \nu}$ is characterized by the classical non-linear solution of eq.~\eqref{eq:K_eq1}. 
To better see the connection between stability and background evolution, it is useful to invert the second relation in \eqref{eq:lagpert} and express ${\cal \tilde K}_{\mu \nu }$. We obtain
\begin{equation}\label{eq:KfunctionofZ}
\mathcal{\tilde{K}}_{\mu \nu} = -\frac{1}{2}\left(Z_{\mu \nu} - \frac{1}{3} Z \eta_{\mu \nu} \right) - \frac{1}{4}\bar \eta_{\mu \nu} + \frac{1}{12}(1+3 c_s^2)\eta_{\mu \nu} \;.
\end{equation}
Using this expression to replace ${\cal \tilde K}_{\mu \nu }$, the equation for the background, eq.~\eqref{eq:K_eq2}, becomes an equation containing only quadratic terms in $Z_{\mu \nu}$, i.e.,
\begin{equation}\label{eq:Zus}
\frac{1}{3}  Z^2 - (Z_{\mu \nu } )^2 =  \frac{3 c_s^2 -1}{6} \;.
\end{equation}
Remarkably,  the terms containing $\Gamma_{\mu \nu}$  have cancelled out: we obtain the same equation as the one derived in \cite{Nicolis:2004qq}  without GWs, although here we have neglected the presence of matter sources, and we have considered a generic $c_s^2$.  This is to be expected since $\dot \gamma_{\mu \nu}$ can be set to zero \emph{locally} by a proper change of coordinates; thus, its value cannot affect eq.~\eqref{eq:Zus}. On the other hand, the solution for $\hat \pi$ requires a global knowledge of the GWs.
We can now use this equation to discuss the stability of the solution.

\subsection{Stability in the absence of GWs}
\label{sec:NicolisRattazzi}
To warm up, we will first review the argument for the case $\Gamma_{\mu \nu}= 0$ and $c_s^2=1$, analogous to the DGP case discussed in \cite{Nicolis:2004qq}. 
A  configuration that turns off at spatial infinity, i.e.~for which $\mathcal{K}_{\mu \nu} = 0$ and $Z_{\mu \nu} = -\eta_{\mu \nu} /2$, is stable in this limit. 
One can show that such a solution cannot become unstable at any other point ${\vect x}$. The proof is made by further assuming that the matrix $Z_{\mu \nu}(x)$ is diagonalizable by means of a Lorentz boost, in such a way that it can be taken to the form $Z^\mu_{\;\nu} = \text{diag}(z_0, z_1, z_2, z_3)$. 

Using this form, eq.~\eqref{eq:Zus} reduces, for  $c_s=1$, to 
 \begin{equation}\label{eq:NicZdiag}
 -\frac{2}{3} \left[ (z_0^2 + \hdots + z_3^2) - (z_0 z_1 + z_0 z_2 + \hdots + z_2 z_3) \right] = \frac{1}{3}\;.
 \end{equation}
In this frame,  stability requires that $z_{\mu}< 0$, for all $\mu = \{ 0, 1,2, 3\}$. Marginally stable solutions, on the other hand, lie on the hyper-planes defined by $z_\mu = 0$, for some $\mu$. A stable solution can become unstable if and only if the solution crosses one of these critical hyper-planes at some intermediate point in the evolution. For this to happen, these critical hyper-planes should intersect the space of solutions. Without loss of generality we can consider the plane $z_0 = 0$, $z_i\neq 0$ in \eqref{eq:NicZdiag}, that now reduces to
\begin{equation}
 - \frac{1}{3} \left[ (z_1 - z_2) ^2 + (z_1 - z_3)^2 + (z_2 - z_3)^2 \right] = \frac{1}{3}\;.
 \end{equation} 
 This equation does not admit any solution because the two sides have different signs: a stable solution at infinity remains stable everywhere. Notice that the right-hand side of the equation above is replaced by $(3 c_s^2 -1)/6$ for a general $c_s$, see eq.~\eqref{eq:Zus}. Therefore the stability of the system, even in the absence of GWs, is not guaranteed for $c_s < 1/\sqrt{3}$. 
 
In the next two sections we are going to explicitly show the presence of instabilities around a GW background, respectively for $c_s < 1$ and $c_s = 1$.  For $c_s > 1/\sqrt{3}$, this is somewhat surprising, since eq.~\eqref{eq:Zus} is qualitatively the same as in the absence of GWs. The catch is that the matrix $Z_{\mu\nu}$ will not be diagonalizable. Indeed, diagonalizability can be proven in the case of non-relativistic sources, but it does not hold for a GW background, which is clearly relativistic.

\subsection{The effect of GWs, $c_s < 1$}\label{sec:csless1}

To study the case $c_s < 1$ we start from eq.~\eqref{eq:stab1}. For large GW amplitudes ($\beta > 1$), the $\gamma \pi \pi$ interaction leads to a wrong sign of the spatial kinetic term for $\delta \pi$. 
However, to confirm this assessment we need to take into account the effect of the tadpole $\gamma \gamma \pi$ and of the self-interactions of $\pi$. The tadpole will generate a background for $\hat \pi$ that, in turn, modifies the action for fluctuations through eq.~\eqref{eq:lagpert}. 

\begin{figure}[t]
\centering
\includegraphics[width=0.5\linewidth]{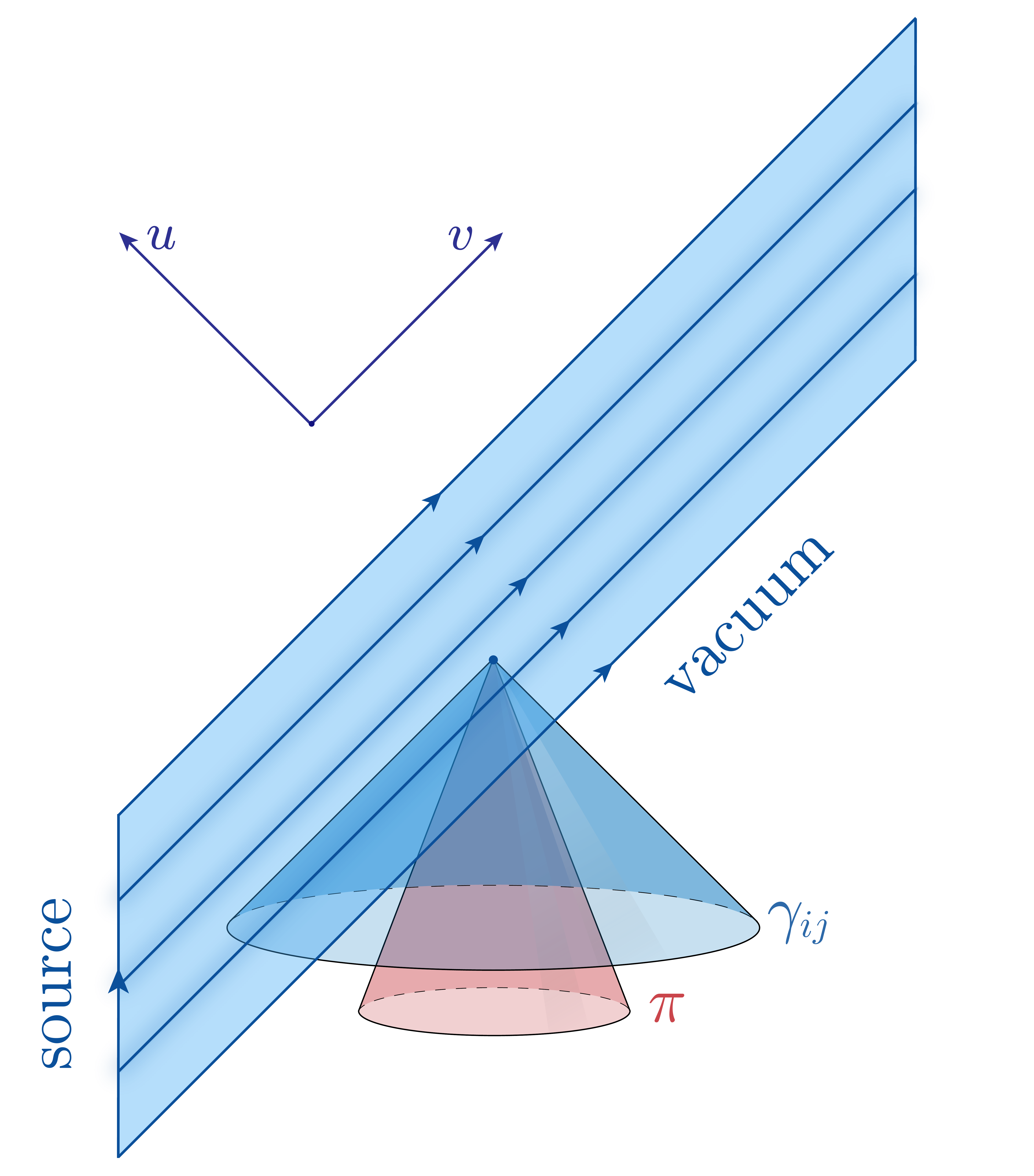}
\caption{:~The GW background we consider is a wavepacket $\gamma_{ij}(u)$. In the case $c_s <1$ the produced $\hat \pi$ is not sensitive to the source of GWs, so that it does not depend on $v$: $\hat\pi(u)$.}  
\label{fig:stripe}
\end{figure}

First, it is convenient to introduce the null coordinates \cite{Creminelli:2019nok}, see Fig.~\ref{fig:stripe},
 \begin{equation}\label{eq:UandScs}
u \equiv t-z \;, \qquad v \equiv  t+z \;, 
\end{equation}
which implies $\partial_t = \partial_u + \partial_v$ and $\partial_z = \partial_v - \partial_u$.
Moreover, the following relations will be useful below,
\begin{align}\label{eq:derivatives}
\partial_t^2 = \partial_u^2 + \partial_v^2 + 2 \partial_u \partial_v\;, \qquad
\partial_z^2 = \partial_u^2 + \partial_v^2 - 2 \partial_u \partial_v\;, \qquad
\partial_t \partial_z = \partial_v^2 - \partial_u^2\;.
\end{align}
In the presence of a background for the GW, of the form $\gamma_{ij} (u)$, we can solve the equation \eqref{eq:stab1} for $\hat \pi$.
In this case, since $c_s <1$, there is no intersection between the region where the source is active and the past light-cone of $\pi$ is finite (see \cite{Creminelli:2019nok}). Therefore we have translational invariance along $v$ (at least as long as we are considering points far away from the emission of $\gamma$). For this reason we will look for solutions of the form $\hat\pi (u)$. 
Notice that the non-linear interaction arising from the cubic Galileon vanishes when $\pi$ depends solely on the variable $u$.
Indeed, eq.~\eqref{eq:derivatives} implies
\be
\partial_t^2 \hat \pi  = \partial_u^2 \hat \pi \,,\qquad
\partial_z^2\hat  \pi  = \partial_u^2 \hat \pi \,,  \qquad \partial_t\partial_z \hat \pi  = -\partial_u^2 \hat \pi \,,
\ee
and thus that 
\begin{align}\label{eq:derivativesCs}
\bar \square \hat \pi = - (1-c_s^2) \partial_u^2 \hat \pi \,, \qquad
(\partial_\mu \partial_\nu \hat \pi )^2 - \square\hat  \pi^2 = 0\, .
\end{align}
Therefore, defining 
\be
\varphi  \equiv  \frac{\hat \pi}{\Lambda_{\rm B}^3}\;,
\ee 
where $\Lambda_{\rm B}^3$ is defined in eq.~\eqref{Mtre}, eq.~\eqref{eq:stab1} gives 
\begin{equation}\label{eq:solXcs}
 \varphi''(u)  =  - \frac{\Gamma_{\mu \nu}^2}{2 (1-c_s^2)} = - \frac{\beta^2 c_s^4}{4 (1-c_s^2)} \cos^2 \left( \omega u \right) \;,
\end{equation}
where we used eq.~\eqref{Gamma}. The solution implies that $\varphi''(u)\leq  0$.
Note that, as one expects, the limit $c_s \rightarrow 1$ is singular: the past light-cone of $\pi$ becomes sensitive to the details of the emission of the GW.

We can now use this solution to compute the kinetic matrix for the $\pi$ fluctuations, eq.~\eqref{eq:lagpert}. Its nonvanishing elements are given  by
\begin{align}\label{eq:ZcompCs}
&Z^{00} = \frac{1}{2} + 2  \varphi''(u) = \frac12 \left[1 - \frac{\beta^2 c_s^4}{1-c_s^2 } \cos^2 (\omega u)  \right] \;,\nonumber \\
&Z^{11} =  - \frac{1}{2} c_s^2+ \Gamma^{11} = - \frac{c_s^2}2  \left[ 1 - \beta \cos (\omega u) \right] \;, \nonumber\\
&Z^{22} =  - \frac{1}{2} c_s^2+ \Gamma^{22} = - \frac{c_s^2}2  \left[ 1 + \beta \cos (\omega u) \right]\;, \\
&Z^{33} = - \frac{1}{2} c_s^2+ 2  \varphi''(u) =  - \frac{c_s^2}2  \left[ 1 + \frac{\beta^2 c_s^2}{1-c_s^2 } \cos^2 (\omega u)  \right] \;, \nonumber\\
&Z^{03} = Z^{30} =  2 \varphi''(u) = - \frac{\beta^2 c_s^4}{2(1-c_s^2) } \cos^2 (\omega u)  \;.\nonumber
\end{align}
The background $\hat \pi$ does not affect the entries $Z^{11}$ and $Z^{22}$ (it contributes only through $\square \hat\pi$ which vanishes since $\hat\pi = \hat\pi(u)$): they feature a gradient instability for $\beta >1$. On the other hand, one can easily verify that the condition $(Z^{03})^2 - Z^{00}Z^{33} > 0$ is satisfied, i.e.~the gradient instability does not appear in this direction. One has a ghost instability, $Z^{00} <0$, for
\be
\label{noghost}
\frac{\beta^2 c_s^4}{1-c_s^2}  > 1 \;.  
\ee

These results seem to contradict what we discussed in the previous section, where we stated that the stability is guaranteed provided $c_s^2 >1/3$. However in order to prove stability one has to assume that the matrix $Z^{\mu \nu}$  is diagonalizable via a boost at each point. This is possible only when $|Z^{03}|  < \frac{1}{2}|Z^{00} + Z^{33}| $ \cite{Dubovsky:2005xd}. In our case this condition gives
\begin{equation}\label{eq:CondDiagCs}
\frac{\beta^2 c_s^4}{1-c_s^2}  < \frac{1-c_s^2}{4} \;.
\end{equation}
Comparing this inequality with eq.~\eqref{noghost}, one sees that the $Z^{\mu \nu}$ is not diagonalizable when there is a ghost instability, so there is no contradiction with the result of Sec.~\ref{sec:NicolisRattazzi}. The inequality \eqref{eq:CondDiagCs} can be written as
\be
c_s^2 < \frac{1}{1+ 2 \beta} \;.
\ee
In the presence of a gradient instability, $\beta >1$, the right-hand side is smaller than $1/3$. Therefore the matrix $Z^{\mu \nu}$ can be diagonalized only if $c_s^2 < 1/3$. Again there is no contradiction with what we discussed above since for $c_s^2 < 1/3$ there is no guarantee of stability.


\subsection{The effect of GWs,  $c_s = 1$}
\label{sec:csone}

Let us now turn to $c_s = 1$. This case is qualitatively different since the light-cone of $\pi$ is as wide as the one of the GWs, see Fig.~\ref{fig:stripe}, so that we should not expect the same kind of instabilities. As before, we take a GW of the form $\gamma_{ij}(u)$, but now one cannot assume that $\hat\pi$ only depends on $u$; in general it will also depend on $v$ and it will be sensitive to the source of GWs. However, for simplicity, we stick to the equations in the absence of sources.

To find $\hat \pi$ we write \eqref{eq:stab1} in terms of the null coordinates $u$ and $v$ and using eq.~\eqref{eq:derivatives} we find
\begin{align}\label{eq:piOperators}
\square \hat \pi &= - 4 \partial_u \partial_v\hat \pi\;, \\
(\partial_\mu \partial_\nu \hat \pi)^2 & = (\partial_t^2 \hat \pi)^2 - 2 (\partial_t \partial_z \hat \pi)^2 +(\partial_z^2 \hat \pi)^2  = 8 \left[ \partial_u^2 \hat \pi \partial_v^2 \hat\pi +(\partial_{u}\partial_v \hat \pi)^2  \right]\;.
\end{align}
Equation \eqref{eq:stab1}, written in terms of $\varphi = \hat \pi \Lambda_{\rm B}^{-3}$, becomes 
\begin{equation}\label{eq:varphi}
\partial_u \partial_v \varphi +4 \left[  \partial_u^2 \varphi \, \partial_v^2 \varphi  - (\partial_u  \partial_v \varphi)^2\right] = -\frac{\Gamma_{\mu\nu}^2}{8}\;.
\end{equation}
(Notice that the coupling $\gamma\pi\pi$ does not contribute.)

It is not clear how to determine the most general solution of the above non-linear equation. However, two solutions can be easily obtained by considering $\varphi (u,v) = U(u) V(v)$. Then, one can make the LHS of \eqref{eq:varphi} independent of $v$ by taking $V(v) = v$.  In this case, the equation takes a very simple form  in terms of $U(u)$,
\begin{equation}\label{eq:separationvariable}
U'(u) ^2 - \frac{1}{4} U'(u) - \frac{\Gamma_{\mu \nu}^2}{32} = 0\;,
\end{equation}
with solutions
\begin{equation}\label{eq:solX}
U'_{\pm}(u) = \frac{1}{8}\left( 1 \pm \sqrt{1+ 2 \Gamma_{\mu \nu}^2} \right) \;,
\end{equation}
where the solution that recovers the linear one at small couplings is $U_-'(u)$. (The conclusions about stability are not altered by considering the other branch.)

We can now check whether the solution \eqref{eq:solX} is stable or not. The kinetic matrix eq.~\eqref{eq:lagpert} is given by
\begin{align}
Z^{00} &= \frac{1}{2} + 2\left[v U''(u) - 2 U'(u) \right]\;, \nonumber \\
Z^{11} &=  - \frac{1}{2} + \Gamma^{11}  + 8 U'(u) \;,  \nonumber \\
Z^{22} &=  - \frac{1}{2} + \Gamma^{22}  + 8 U'(u) \;, \label{eq:Zcomp} \\
Z^{33} &= - \frac{1}{2} + 2\left[v U''(u) + 2 U'(u) \right]\;, \nonumber\\
Z^{03} &= Z^{30} =  2 v U''(u)  \nonumber\;.
\end{align}
First, we focus on possible gradient instabilities. One can easily check, using \eqref{eq:solX}, that the components $Z^{11}$ and $Z^{22}$ are negative, so there is no gradient instability in these directions. The matrix is non-diagonal in the block $t$-$z$ and stability requires $(Z^{03})^2 - Z^{00}Z^{33} > 0$. In our case
\begin{align}\label{gradient:condition1}
	(Z^{03})^2 - Z^{00}Z^{33} = \left( \frac{1}{2} - 4 U'(u) \right)^2 \geq 0 \;;
\end{align}
the matrix $Z^{\mu\nu}$ is thus free from gradient instabilities. 

Let us turn now to ghost instabilities. As already pointed out at the beginning of Sec.~\ref{sec:stabArg}, ghosts are present whenever $Z^{00}$ becomes negative.
From \eqn{eq:Zcomp} we see that this is possible: the term linear in $v$ can be negative and larger than the other positive contributions. 
To see this more explicitly, we can replace $U(u)$ in $Z^{00}$ with the solution \eqref{eq:solX}. We get
\begin{equation}
Z^{00} = \frac{1 +  2 \Gamma_{\mu \nu}^2  - v \Gamma_{\mu \nu}\partial_u \Gamma^{\mu \nu}}{2 (1+ 2 \Gamma_{\mu \nu}^2)^{1/2}} \simeq \frac{1 -  \Gamma_{\mu \nu}^2  \omega v   }{2 (1+ 2 \Gamma_{\mu \nu}^2)^{1/2}}\;,
\end{equation}
where in the last equality we used that $\omega v \gg 1$ and we approximated $\partial_u \Gamma_{\mu \nu} = \omega \Gamma_{\mu \nu} \tan ( \omega u) \simeq \omega \Gamma_{\mu \nu}$
 (valid for a plane wave). 
Using $\Gamma_{\mu \nu}^2 \simeq \beta^2/2 $, the condition to avoid a ghost becomes
\be
\beta^2 \lesssim \frac{2}{\omega v } \;.
\ee
Since after a few oscillations $\omega v \gg 1$, we conclude that also for $c_s = 1$ the system becomes unstable. Notice that also in this case the matrix $Z^{\mu\nu}$ is not diagonalizable (the condition $|Z^{03}|  < \frac{1}{2}|Z^{00} + Z^{33}| $ is not satisfied by eq.~\eqref{eq:Zcomp}, but the two sides are actually equal) so that there is no contradiction with the result of section \ref{sec:NicolisRattazzi}.

Even if the solution we studied is not unique and does not take into account the effect on $\hat\pi$ of sources, we conclude that the system is generically unstable and avoids the stability argument of \cite{Nicolis:2004qq}, because the GW background is relativistic and gives a non-diagonalizable $Z^{\mu\nu}$ (\footnote{Even in the absence of GWs, perturbations around a plane wave $\hat\pi(u)$, with $c_s=1$, are unstable \cite{Burrage:2011cr}, as it is easy to check. This suggests that the instability is generic in a relativistic setting.}). For the same reason, one expects the system to be unstable also for $c_s>1$ (although it is not clear whether a theory of this kind allows a standard UV completion). In this case, it is not clear how to find a simple ansatz for the solution $\varphi(u,v)$, so that a dedicated study is left to future work.

\subsection{Vainshtein effect on the instability}
So far we assumed that GWs are the only source of $\hat \pi$. However, astrophysical objects also source $\hat \pi$ and a corresponding matrix $Z^{\mu\nu}$. As discussed above, in the presence of non-relativistic sources this matrix is healthy and it gives rise to the so-called Vainshtein effect: a large $Z^{\mu\nu}$ gives a more weakly coupled theory, in which the effect of $\pi$ is suppressed.  The Vainshtein effect will also suppress the instability we are studying: the astrophysical background makes the kinetic term large and healthy, while the dangerous vertex $\gamma\pi\pi$ is {\em not} enhanced (one does not have a term $\partial^2\hat\pi \, \dot\gamma \partial \pi \partial\pi$, since it would have too many derivatives). Therefore, in regions with large $Z^{\mu\nu}$ the parameter $\beta$ is effectively suppressed and the  instabilities can thus be stopped. However, the condition of large $Z^{\mu\nu}$ cannot be maintained over cosmological scales. Both analytical arguments \cite{Cusin:2017mzw} and simulations \cite{Schmidt:2009sg} indicate that Vainshtein screening is negligible over sufficiently large scales, say larger than 1 Mpc; see Sec.~\ref{sec:constraints} for a more detailed discussion. This means that averaged over these large scales the effect of astrophysical sources is negligible\footnote{Since astrophysical sources are with good approximation non-relativistic, the entries $Z^{0i}$ of the matrix $Z^{\mu\nu}$ are negligible (see discussion at the beginning of Sec.~\ref{sec:stabArg}). In order to stabilize the gradient instability one should have that all the eigenvalues of the spatial part of the matrix $Z^{ij}$ are large, much larger than the standard kinetic term, i.e.~parametrically larger than unity (in absolute value). This means that also the trace should be parametrically larger than unity. To avoid the instability these conditions should be maintained over all the trajectory of the GW, i.e.~over cosmological distances. This however cannot happen. If the trace were large over large regions, it would imply that the trace of the average of $Z^{\mu\nu}$ over a large region is sizeable. This is in contradiction with the statement that linear perturbation theory is recovered over sufficiently large scales. }. Since the GWs we observe travel over cosmological distances, one expects that on average the effect of Vainshtein screening is small and that over most of their travel the gradient instability is active. We will come back to this point below, in Sec.~\ref{sec:constraints}.


\section{Fate of the instability}
\label{sec:fateofin}
In order to understand the implications of the instability we discussed, one would like to know the fate of it. In this Section we want to argue that the dynamics of the instability and its endpoint are UV sensitive and cannot be studied without knowing the UV theory. First of all, notice that it is not possible to follow the development of the instability looking at what happens at the matrix $Z^{\mu\nu}$ in the presence of the growing perturbations. Since the most unstable modes are the shortest, the instability generates a configuration of $\hat\pi$ with very large gradients and the analysis of the previous sections is only useful to understand the behaviour of modes with wavelength much shorter than the variation of the background. 

Since we do not know of any UV completion of the theories we are discussing, to gain some intuition on the possible outcome of the instability we now discuss a toy model that features gradient and ghost-like instabilities, and whose UV completion is known. Consider a $U(1)$-symmetric theory for a complex scalar $\phih$, with a quartic mexican-hat potential, in the absence of gravity\footnote{This analysis is based on unpublished work with A.~Nicolis. See also \cite{Babichev:2018twg}.}:
\be \label{L_UV}
{\mathscr L}_{\rm UV} = - |\partial \phih|^2 - V (|\phih|) \; , \qquad  V (|\phih|) = \lambda \big( |\phih|^2 -v^2 \big)^2 \;.
\ee 
In the broken phase with $\langle \phih \rangle = v$ we have a massless degree of freedom (the Goldstone boson), and a heavy one (the `Higgs'). It makes sense to integrate out the latter and write down a low-energy effective theory for the former.

For small $\lambda$, one can integrate out the Higgs at tree level. Let us define $\phih = \phih_0 \, \exp{(i \pih)}$. If we are interested in terms with the minimum number of derivatives acting on $\pih$, one can solve the classical equation of motion for a constant $\phih_0$ in a constant $X$ field, $X \equiv (\partial\pih)^2$, and plug the result back into the action.
One gets
\be \label{barphi}
\phih_0^2 = - \frac{1}{2\lambda} X + v^2 = \frac{1}{4 \lambda}(\mu^2 -2 X)\; ,
\ee
where $\mu^2 \equiv 4 \lambda v^2$ is the mass of the radial direction.
Plugging this back into the action, one gets the Lagrangian 
\be\label{X2}
P(X) \simeq - \frac{1} {4 \lambda} \,  X \big( \mu^2 - X \big) \;.
\ee
Remarkably, the tree-level effective action stops at quadratic order in $X$, that is at fourth order in $\pih$. The function $P(X)$ will receive corrections suppressed by $\lambda$ at loop level. Notice that the validity of this action is not limited to small $X$, provided {\em derivatives} of $X$ are small: operators with derivatives acting on $X$ are suppressed by powers of $\partial/\mu$. 

Consider a background $\hat\pih$ with $\partial_\mu \hat\pih \equiv C_\mu$ and small perturbations about it, $\hat\pih + \delta\pih$. The matrix $Z_{\mu\nu}$, see eq.~\eqref{eq:lagpert}, is given by
\be
Z^{\mu\nu} = 2 \hat P'' C^\mu C^\nu + \hat P' \eta^{\mu\nu} \;.
\ee
If $C^\mu$ is time-like, that is if $\hat X < 0$, we can choose a frame such that $C^0 = \pm \sqrt{-\hat X}$, $\vec C = 0$. In this frame we have
\be
{\mathscr L}_2 = -\big( 2 \hat P'' \hat X + \hat P' \big) \delta \dot \pih^2 +  \hat P' (\nabla \delta \pih)^2 \;, \qquad \hat X < 0 \;.
\ee
For stability we thus want 
\be \label{stability}
 2 \hat P'' \hat X + \hat P' < 0 \; , \qquad \hat P' < 0 \; .
\ee
If instead $C^\mu$ is space-like, $\hat X > 0$, we can go to a frame where $C^0 =0$, $ |\vec C| = \sqrt{\hat X}$, where we get
\be \label{L2_negativeX}
{\mathscr L}_2 = - \hat P' \delta \dot \pih^2 +  \hat P' (\nabla_\perp \delta\pih)^2 + \big( 2 \hat P'' \hat X + \hat P' \big) (\nabla_\parallel \delta\pih)^2 \;, \qquad \hat X > 0 \; .
\ee
The parallel and normal directions are of course relative to $\vec C$. We thus see that in this case too the conditions for stability are those given in eq.~(\ref{stability}).

For the case we are studying, eq.~\eqref{X2}, one has
\be
2 \hat P'' \hat X + \hat P' = \frac 1 {4\lambda} (6 \hat X - \mu^2) \; ,\qquad \hat P' = \frac 1 {4\lambda} (2 \hat X - \mu^2) \;.
\ee
The system is stable for 
\be
\hat X < \frac16{\mu^2} \; .
\ee
It is interesting that for such values of $\hat X$, the propagation speed is always subluminal---a non-trivial check about the consistency of the effective theory. 
In the range
\be
\frac{1}{6} \mu^2 <  \hat X < \frac1{2} \mu^2 \;, 
\ee
the $(\nabla_\parallel \delta\pih)^2$ in eq.~(\ref{L2_negativeX}) has the wrong sign, thus signaling a tachyon-like instability which, unlike a real tachyon instability, is dominated by the UV. That is, we have exponentially growing modes  $\sim e^{k_\parallel t}$. The shorter the wavelength, the faster the growing rate. Finally, for 
\be
\hat X > \frac1{2} \mu^2 
\ee
all terms in  eq.~(\ref{L2_negativeX}) have wrong signs. This in the low-energy effective theory looks like a ghost-like instability.

It is interesting to understand these pathologies in terms of the UV theory (\ref{L_UV}). There, the kinetic energy is positive definite. There is no room  for ghost-instabilities, and the only instabilities present in certain regions of field space are real tachyons, with a decay rate of order $\mu$. 
Let us therefore consider small fluctuations of the radial mode $\phih_0$ and of $\pih$ in the UV theory, about a background configuration with constant $\hat X$ and  $\hat \phih_0$, related by eq.~(\ref{barphi}),
\be
\phih_0 \to \hat\phih_0 + \varphih \; , \qquad \pih \to \hat\pih + \delta\pih \; .
\ee
Expanding the Lagrangian (\ref{L_UV}) at quadratic order we get
\be \label{L_UV_fluctuations}
{\mathscr L}_{{\rm UV}} \to  \dot \varphih^2 + \delta\dot {\tilde \pih} ^2 - \Big( \begin{array}{c} \varphih \\ \delta\tilde \pih  \end{array}   \Big)  \cdot 
 \Big( \begin{array}{cc}  -\nabla^2 + (-2 \hat X+\mu^2) & 2 \sqrt{\hat X} \, \nabla_\parallel \\ 
- 2 \sqrt{\hat X} \, \nabla_\parallel & -\nabla^2
 \end{array}   \Big) \cdot \Big( \begin{array}{c} \varphih \\ \delta\tilde \pih  \end{array}   \Big) \;,
\ee
where we canonically normalized the angular fluctuations by defining $\delta\tilde \pih = \hat \phih_0 \, \delta\pih$, and we specialized to the positive-$\hat X$ case (spacelike $C^\mu$), given that this is the region where the pathologies discussed above show up.

First, notice that for $\hat X = \frac12 \mu^2$ the mass term for the radial fluctuation $\varphih$ goes to zero. This means that at this particular point in field space we cannot get a local low-energy effective theory for the $\tilde \pih$ by integrating $\varphih$ out. Also, at the same point the radial background $\hat \phih_0$ goes to zero---see eq.~(\ref{barphi})---and it remains zero for even larger values of $\hat X$. We thus see that the ghost instability we encounter in the low-energy theory for the angular mode starting from $\hat X = \frac12 \mu^2$, is a sign that at those values of $\hat X$ the low-energy theory just makes no sense---the derivative expansion breaks down at zero energy.

Then, we see from the structure of eq.~(\ref{L_UV_fluctuations}) that the background configuration is stable if and only if the gradient/mass matrix has positive eigenvalues. For plane-waves with momentum $\vec k$ parallel to $\vec C = \vec \nabla \hat\pih$, the determinant of such matrix is
\be
k^2_\parallel \, \big[ k^2_\parallel + (\mu^2- 6 \hat X)\big] \; .
\ee
We thus see that for $\hat X > \frac16 \mu^2$, the gradient/mass matrix develops a negative eigenvalue in a finite range of momenta,  $0 <k^2_\parallel < (6 \hat X - \mu^2)$. This signals an instability with a rate of order $\mu$. Indeed from the low-energy viewpoint the instability was UV-dominated, and we see that in the UV theory it is saturated at $k_\parallel \sim \mu$. At higher energies the UV theory makes perfect sense.

What can we learn from this example about the instability induced by GWs? 
\begin{itemize}
\item Instabilities can arise from a perfectly sensible theory when one goes in a certain region of field space and from the EFT perspective one can only conclude that the instability exists in the regime of validity of the EFT itself: the theory may be completely healthy in the UV. 

\item The example we discussed has a well-defined Hamiltonian bounded from below, hence at most the instability can convert this finite amount of energy into the unstable modes. Therefore one can only conclude that an energy of order $\Lambda_{\rm UV}^4$, with $\Lambda_{\rm UV}$ the cut-off of the theory, is damped into the unstable modes; all further developments depend on the UV completion. Since in our case $\Lambda_{\rm UV}^4$ is parametrically smaller than the energy density of the GWs (which accidentally is of order $\Lambda_2^4$ for the typical amplitudes and frequencies detected by LIGO-Virgo) one cannot conclude that the GW signal will be affected. 
\item The appearance of the instability may signal that the EFT breaks down. This happens in the example above in the case of the ghost instability: the range of applicability of the EFT shrinks to zero. The regime of validity of the EFT is not only determined by the requirement that frequencies are sufficiently small, but it can be modified in the presence of a sizeable background.  Therefore it may be that the instability we studied is simply telling us that the EFT of DE breaks down. This means that we are unable to describe the propagation of GWs unless we know the UV completion of the theory.

\end{itemize}

Notice that both in the case in which the instability can be described within the EFT and in the case in which the EFT breaks down at the instability, in order to continue the time evolution of the system one needs the UV completion.


\section{Phenomenological consequences}
\label{sec:constraints}

Let us explore the phenomenological consequences of the instability we studied. First of all, as it is clear from the toy model we described in the previous section, without a UV completion one cannot conclude that a sizeable amount of energy goes into $\pi$. The instability may be saturated at the cut-off scale $\Lambda_3$ or even at a lower scale. This means that it is not guaranteed that the instability leads to a backreaction on the GW signal that can be seen at the interferometers\footnote{In fact, using eq.~(4.5) of \cite{Creminelli:2019nok} one can straightforwardly show that $\Delta \gamma/\bar{\gamma} \sim \Lambda_3^4/(\Lambda_2^4h_0^+) \ll 1$, where $\bar{\gamma}$ and $\Delta\gamma$ denote the GW background and its modification respectively.}. 
In the following, we will concentrate on the question of whether a generic point in the Universe is affected by the instability. For this we do not need to focus on the particular events observed by LIGO-Virgo (or eventually LISA  and pulsar timing array, see e.g.~\cite{Hobbs:2009yy}) but one has to consider the effect of all GW emissions. 

 Let us neglect momentarily the Vainshtein effect. The Universe is populated by binary systems and these trigger the instability in points that are close enough to the source to have $\beta >1$. Let us divide the Universe in spheres of $10$~Mpc radius and ask whether the instability is triggered in these regions. Since in first approximation the Universe is homogeneous on scales of $10$~Mpc, one expects that all regions behave approximately in the same way.  If within a region and in a time comparable to the age of the Universe, there is at least one binary event that gives $\beta >1$ at a distance of $10$~Mpc, one can conclude that this event will trigger the instability over the whole sphere (and thus in the whole Universe). In the following we are also going to explore regions of $1$~Mpc. In this case, since the Universe is inhomogeneous on this scale, using the same criteria as before one can only conclude that sufficiently dense regions reached the instability. Indeed the events will be mostly localized in overdensities and may not be able to trigger the instability in underdense regions.  
 
 The parameters needed to characterize the instabilities discussed in Sec.~\ref{sec:csless1} are the amplitude $h_0^+$ and the frequency $f$. Long before the merger, the amplitude $h_0^+$ can be written as (see for example \cite{Maggiore:1900zz})
\begin{equation}
\label{eq:hstrain}
h_0^+ \sim \frac{1}{\sqrt{2}} \cdot \frac{4}{r} (G M_c)^{5/3} (\pi f)^{2/3}\;,
\end{equation} 
where $r$ is the distance from the binary,  $M_c$ is the chirp mass and $f$ the GW frequency. (The factor of $1/\sqrt{2}$ comes from our non-standard definition of $h_0^+$.) This is a reasonable approximation until the orbit reaches the innermost stable circular orbit (ISCO)\footnote{In a Schwarzschild geometry the ISCO is located ar $r_{\rm ISCO} = 6 G m$, where $m$ is the total mass of the binary. Assuming equal masses and using Kepler's law to convert into frequency, we find $f_{\rm ISCO} \simeq 0.034 /(\pi G  M_c)$.}. 

\begin{figure}[t]
\centering
\includegraphics[width=0.6\linewidth]{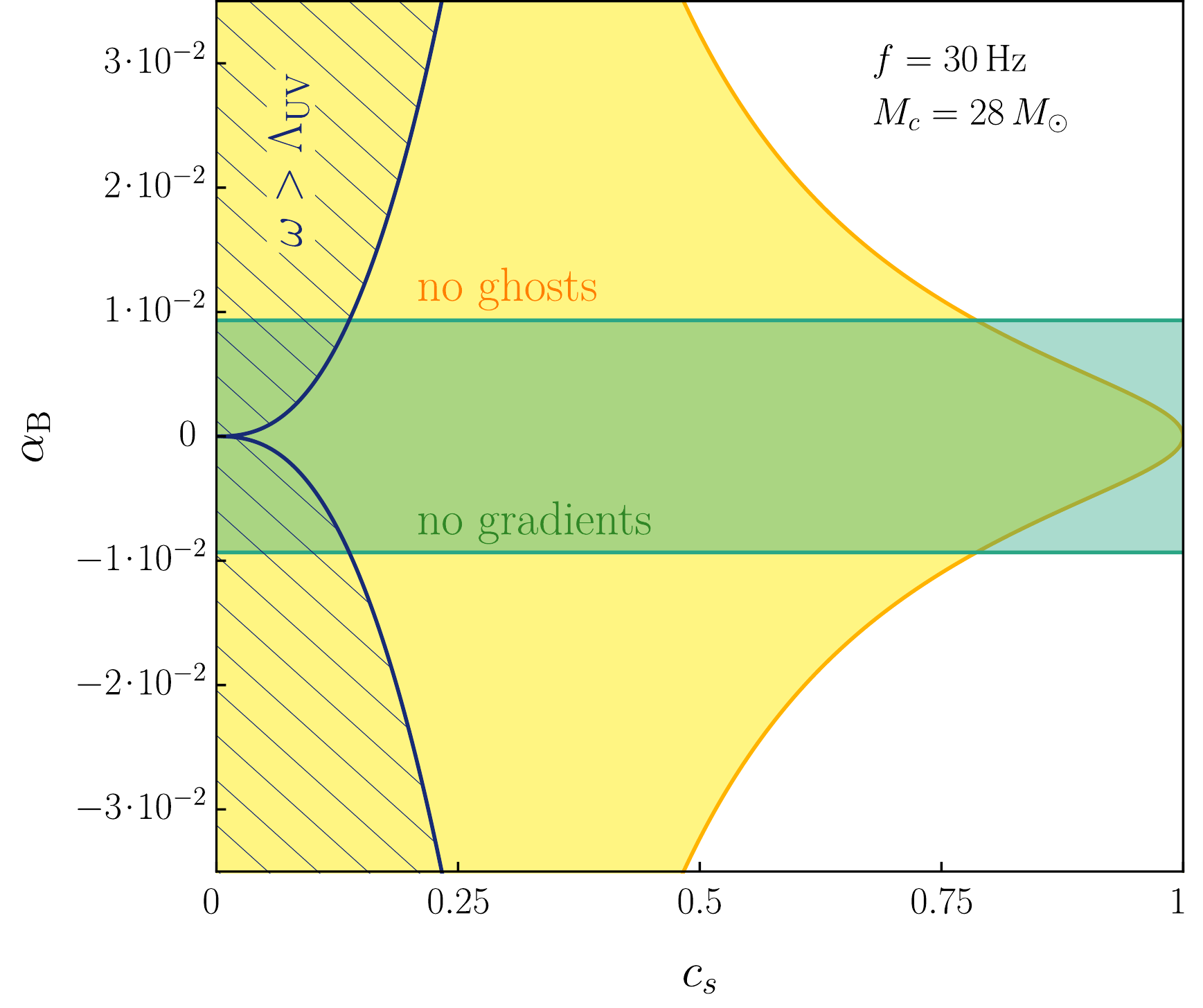}
\caption{:~Stability regions on the plane $(c_s, \alpha_{\rm B})$. The yellow region indicates where ghost instabilities of eq. \eqref{noghost} are absent in which $\beta < c_s^{-2}\sqrt{1-c_s^2}$. The green region indicates where gradient instabilities are absent, i.e.~$\beta < 1$. The fact that this curve is independent of $c_s$ follows from the choice $\alpha = {1}/({2 c_s^2})$, see eqs.~(\ref{InterScale}) and (\ref{beta}). In the  region with the blue diagonal grid, the frequency of the GW is above the perturbative unitarity bound, $\omega > \Lambda_{\rm UV}$ (see footnote \ref{10}), and our analysis cannot be applied. In the plot we have used $M_c = 28 M_\odot $ and $f = 30\, \rm{Hz}$.}  
\label{fig:stellar}
\end{figure}
Figure~\ref{fig:stellar} focusses on stellar mass BHs; for concreteness we chose $M_c = 28 \, M_\odot $ as for GW150914 and $f = 30\, \rm{Hz}$. We take the distance to be $1\, \rm{Mpc}$. Taking a distance of 
$10\, \rm{Mpc}$ would require, in order to keep the same $h_0^+$, to consider times closer to the coalescence. However this corresponds to larger frequencies and one goes in a regime that cannot be trusted, since the frequency is higher than the unitarity cut-off.\footnote{\label{10}The cut-off can be obtained as the energy scale at which perturbative unitarity is lost. In order to explicitly get such scale for $m_3^3$ we focus on the leading term in \eqref{InteractionLagrangian}: the dominant interaction in the small-$c_s$ limit is $\sim - \nabla^2 \pi (\partial_i \pi)^2/{\Lambda_{\rm B}^3}$. Following \cite{Baumann:2011su,Pirtskhalava:2015zwa} we find that for such interaction perturbative unitarity in the $\pi \pi \rightarrow \pi \pi$ scattering is lost  when 
\be
\frac{\omega^6}{\Lambda_{\rm B}^6 c_s^{11}} < \frac{3\pi}{4}\;, 
\ee
where here $\omega$ is the energy of $\pi$.} In Fig.~\ref{fig:stellar} we plot the gradient and ghost instabilities in the plane $(c_s, \alpha_{\rm B})$ together with the unitarity cut-off. Models with $\alpha_{\rm B} \gtrsim 10^{-2}$ are affected by one or both instabilities, but the cut-off is quite close.

\begin{figure}[t]
\centering
\includegraphics[width=0.8\linewidth]{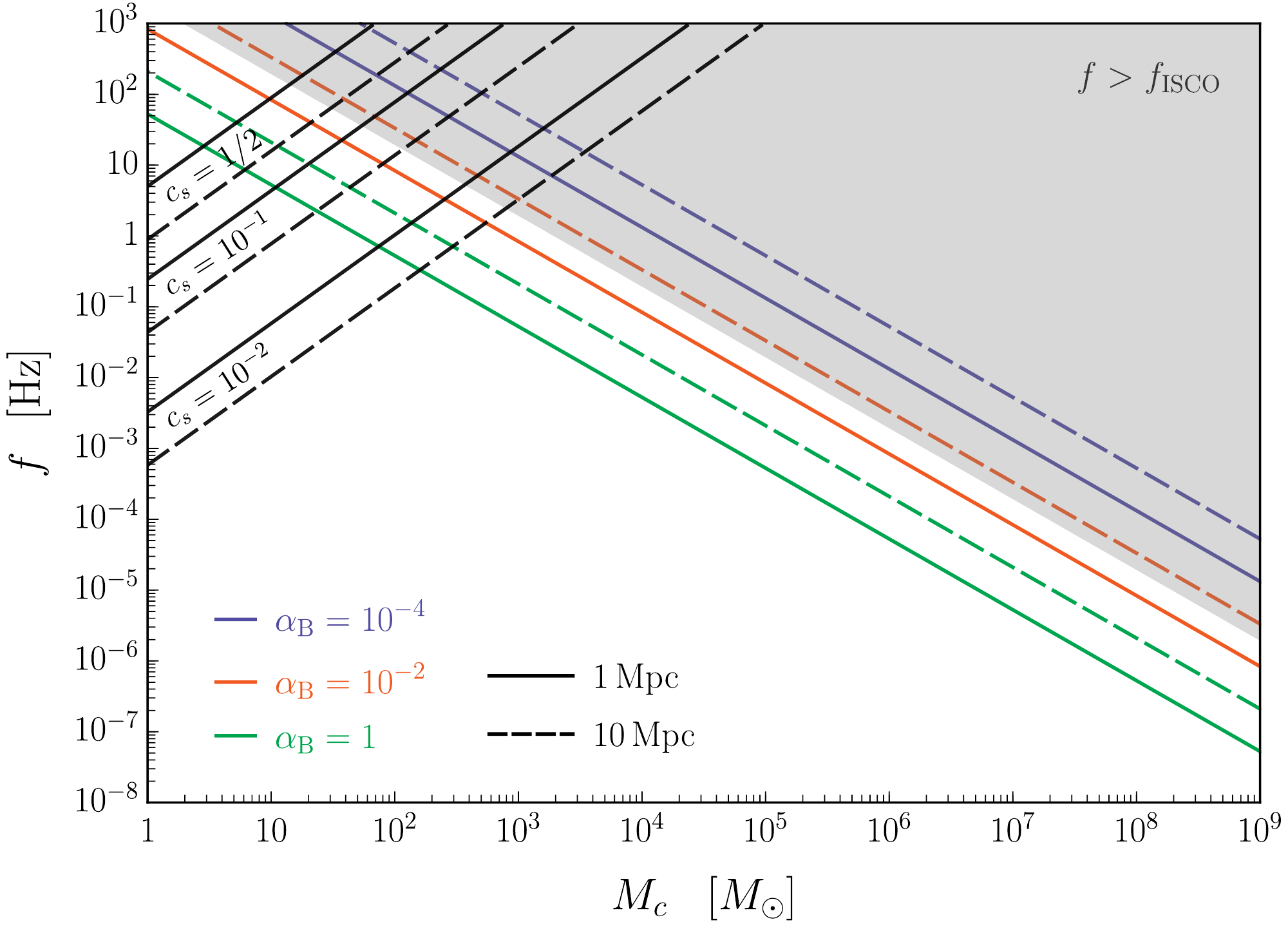}
\caption{:~Gradient-instability lines, $\beta =1$, for different value of $\alpha_\textrm{B}$ as a function of the chirp mass of the binary system, evaluated at a distance of $1\, \textrm{Mpc}$ (solid lines) and $10\, \textrm{Mpc}$ (dashed lines). The grey region cannot be trusted because it would correspond to extrapolating the orbit beyond the ISCO. Regions above the black lines have frequencies larger than the unitarity cut-off $\omega > \Lambda_{\rm UV}$ (the three lines correspond to different values of $c_s$). At fixed $\beta = 1$, we expressed the cut-off frequency as a function of $M_c$ using \eqref{eq:hstrain}. All lines are evaluated with the choice $\alpha = {1}/({2 c_s^{2}})$. }  
\label{fig:MBH}
\end{figure}
On the other hand, if one considers  massive BHs, frequencies are many orders of magnitude smaller than the unitarity cut-off. In Fig.~\ref{fig:MBH} we plot the threshold $\beta =1$ as a function of the chirp mass of the binary for $1\, \rm{Mpc}$ and $10\, \rm{Mpc}$ distances. Independently of the chirp mass, the instability is triggered close to the ISCO for values of $\alpha_{\rm B}$ that are of interest for future LSS experiments, i.e.~$\alpha_{\rm B} \gtrsim 10^{-2}$.
Although there is some degree of uncertainty on the rate of massive BH mergers, one can be quite sure that in a region of $10\, \rm{Mpc}$ many mergers of halos,  and therefore binary mergers of massive BHs, took place in the last Hubble time. To be more quantitative, in the range $10^7 M_{\odot} < M_c < 10^8 M_\odot$ one estimates between 5 and 50 events in a volume of  $10\, \rm{Mpc}$ radius between $z=1$ and $z=0$ \cite{Bonetti:2017lnj}. Rates are larger, but considerably more uncertain, for smaller masses \cite{Bonetti:2018tpf}.

Let us now discuss the role of  screening. As we discussed above, in  regions with large field non-linearities the threshold of instability can be lifted by the Vainshtein mechanism. If the typical radius at which the screening is effective is of order $10\, \rm{Mpc}$ or smaller, then our conclusions do not qualitatively change. There may be very non-linear regions where the instability did not occur, but in most of the Universe the instability takes place.
Following \cite{Cusin:2017mzw}, one can estimate the scale at which the Vainshtein mechanism is relevant assuming a power-law Universe with matter power spectrum $P(k) \propto k^n$, where the relevant value near the non-linear scale for the real Universe is $n \simeq -2 $. In our case one finds $\lambda_{\rm V}  \sim [\alpha_{\rm B}/(c_s^2 \alpha)]^{\frac{4}{3+n} } \lambda_{\rm NL}$, which shows that for small $\alpha_{\rm B}$ the Vainshtein scale $\lambda_{\rm V} $ is in general much shorter than $10\, \rm{Mpc}$, which roughly corresponds to the non-linear scale for structure formation $\lambda_{\rm NL}$ (see also \cite{Schmidt:2009sg} for an estimate of the Vainshtein scale in numerical $N$-body simulations, confirming these estimates).

What can we conclude if a model lies in the unstable region? As we discussed, the endpoint of the instability is unknown and requires knowledge of the UV. Naively one can imagine that a certain amount of $\pi$s with energy close to the cut-off is generated until their backreaction stops the instability. It looks difficult to argue that the theory around this new state will resemble the original one and give similar predictions: the $\pi$s produced by the instability must qualitatively change the theory to make it stable, so that one expects that also the other predictions of the theory will be affected. One cannot make any firm prediction without understanding the fate of the instability and this requires a UV completion. 

Another possibility is that the EFT breaks down at the instability, so that the instability itself cannot be trusted. Notice however that the frequencies involved may be as low as $10^{10}$ km. In this case one has to declare the impossibility to say anything about any process that has to do with GWs. Moreover all the successes of GR on shorter scales cannot be explained. Analyticity arguments can be used to argue that a theory with an approximate Galilean symmetry must break down at a very large scale, of order $10^7$ km in the range of parameters we are discussing \cite{Bellazzini:2019xts}. Although it is not straightforward to apply these arguments in a cosmological context, where Lorentz invariance is spontaneously broken, it is an independent indication that the theories at hand must break down at extremely large scales.


\section{Beyond Horndeski: \texorpdfstring{$\tilde{m}_4^2$}{\tilde m_4^2}-operator}\label{sec:m4}

The analysis of the previous sections focused on the stability of cubic Horndeski theories. Here we want to consider another quadratic operator of the EFT of DE that survives after GW170817 \cite{Creminelli:2017sry}.  The operator 
\begin{equation}
S_{\tilde m_4}  =\int  \text d^4 x   \sqrt{-g} \, \frac{\tilde{m}_4^2 (t)}{2} \delta g^{00} \left({}^{(3)}R + \delta K_\mu^\nu \delta K_\nu ^\mu  - \delta K^2 \right) \;, \label{eq:bh}
\end{equation}
where ${}^{(3)}\! R$ denotes the 3d Ricci scalar of the hypersurfaces at constant $t$, is not constrained by the requirement that GWs travel at the speed of light \cite{Creminelli:2017sry}. However, it is highly constrained by the perturbative and resonant decay $\gamma \rightarrow \pi \pi$, studied in~\cite{Creminelli:2018xsv,Creminelli:2019nok}. The perturbative bound is of the order
\begin{equation}
\label{eq:pertaH}
|\alpha_{\rm H}| \lesssim 10^{-10} \;, \qquad \alpha_{\rm H} \equiv 2 \frac{\tilde m_4^2}{\MP^2} \;.
\end{equation}
 The Lagrangian of $\pi$ in the presence of all the relevant non-linearities schematically reads \cite{Creminelli:2018xsv} (we follow the notation of \cite{Creminelli:2019nok})
\begin{align}
\mathscr{L}_\pi = - \frac{1}{2} \bar{\eta}^{\mu\nu} \partial_\mu \pi \partial_\nu \pi + \frac{1}{\Lambda_\star^3}\ddot{\gamma}_{ij}\partial_i\pi\partial_j\pi - \frac{(\partial\pi)^2}{\Lambda_\star^3}\partial^2\pi + \frac{(\partial \pi)^2}{\Lambda_c^6} [(\Box\pi)^2-(\partial_\mu\partial_\nu\pi)^2] - \frac{\alpha_{\rm H}}{2\sqrt{\alpha}H\MP}\dot{\pi}\dot{\gamma}_{ij}^2  \;,  \label{lagrangian:m4}
\end{align}
where $\Lambda_\star \simeq \alpha_{\rm{H}}^{-1/3}\alpha^{1/3}\Lambda_3$ and $\Lambda_c \simeq \alpha_{\rm{H}}^{-1/6}\alpha^{1/3}\Lambda_3$. The second term gives an instability similar to the one discussed above and we can define, following \cite{Creminelli:2019nok}, 
\be    
\label{betadef2}
\beta \equiv  \frac{2 \omega^2 \MP h_0^+}{ c_s^2 | \Lambda_\star^3 |} = \frac{\sqrt{2} | \alpha_{\rm H}| }{\alpha c_s^2  } \left( \frac{\omega }{H}\right)^2 h_0^+ \;.
\ee

   \begin{figure}[t]
   \centering
   \includegraphics[width=0.8\linewidth]{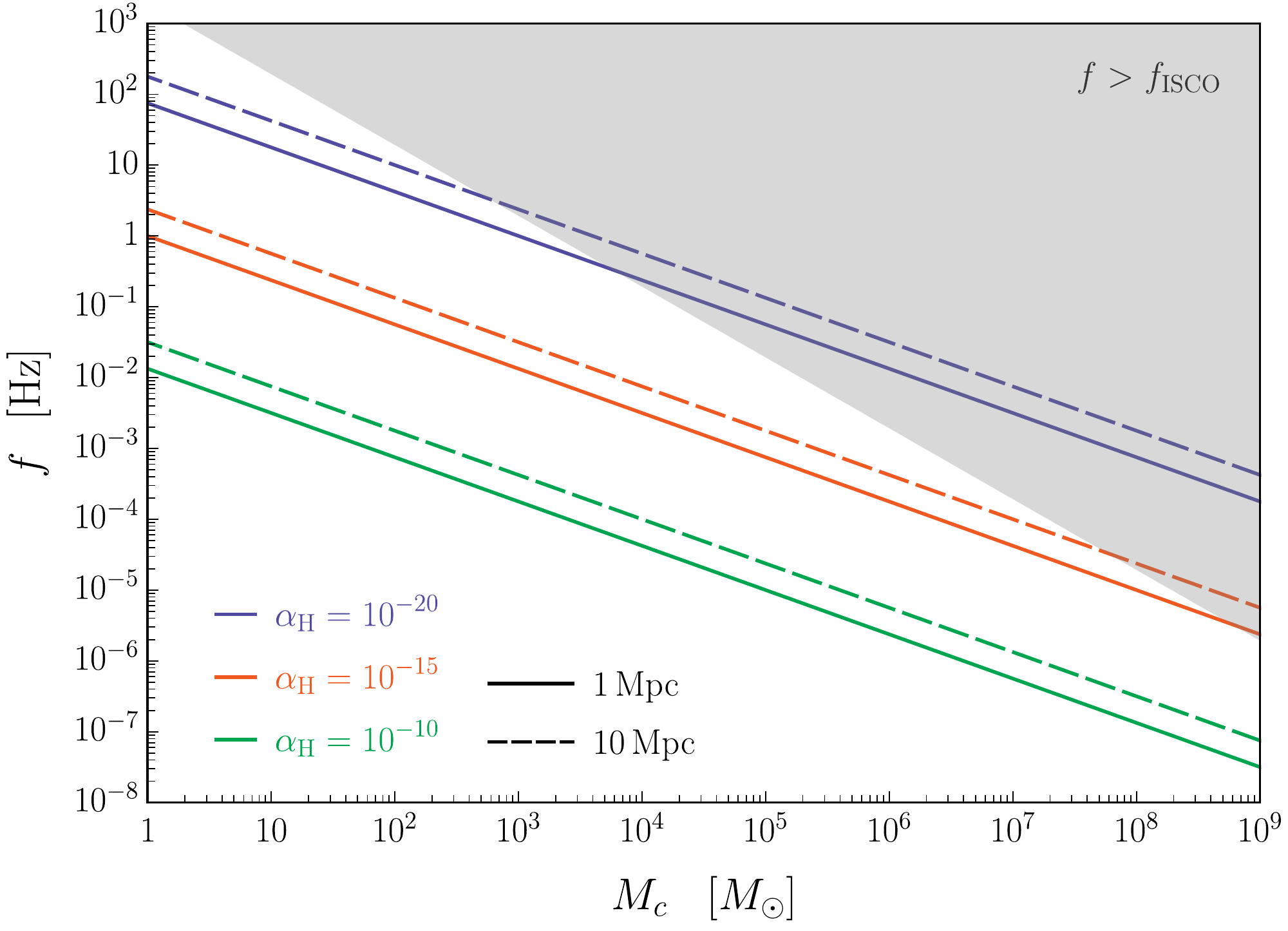}
   \caption{\label{fig:MBH2}:~ Gradient-instability lines $\beta =1$ for
   different value of $\alpha_\textrm{H}$ as a function of the chirp
   mass of the binary system. The grey region cannot be trusted because
   it would correspond to extrapolating the orbit beyond the ISCO.}
   \end{figure} 
It turns out that the analysis for this case is simpler, since to assess the stability of the system it is enough to look at the $\gamma\pi\pi$ interaction, while all additional non-linearities are negligible: the system is unstable for $\beta >1$. We are going to verify this statement below. In Fig.~\ref{fig:MBH2} we plot the instability region as a function of the chirp mass and frequency. Notice that in this case the unitarity cut-off of the theory does not appear in the figure, since it is much higher that the frequencies of interest. The absence of instability is a constraint much tighter than the perturbative bound of eq.~\eqref{eq:pertaH}. On the other hand, the narrow-resonance regime gives even better constraints for $\alpha_{\rm H}$, of the order of $\beta \lesssim 10^{-2}$ (see Fig.~5 of \cite{Creminelli:2019nok}).  

Let us now verify that the other non-linear terms in the Lagrangian of eq.~\eqref{lagrangian:m4} can be neglected. (For simplicity in the following we take $\alpha \sim 1$ and $c_s \sim 1$.) First of all, let us estimate the size of the induced background $\hat\pi$ sourced by $\gamma\gamma\pi$. Neglecting $\pi$ non-linearities and using the Lagrangian (\ref{lagrangian:m4}), one can estimate
\begin{align}
\hat{\pi} \sim \frac{\MP\alpha_{\rm{H}}\omega (h_0^+)^2}{H} \;. \label{piHat}
\end{align}
In order for this estimate to be correct we must check that the cubic and quartic self-interactions of \eqref{lagrangian:m4} are negligible. At the level of the equations of motion, the contributions of the cubic and quartic terms are schematically given by $ \mathcal{E}_{(3)}\sim (\pt^2 \hat{\pi})^2 \, \Lambda_\star^{-3}$ and $ \mathcal{E}_{(4)}\sim (\pt^2 \hat{\pi})^3 \, \Lambda_c^{-6}$. These have to be compared with $\mathcal E _{(2)} \sim \pt^2 \hat{\pi}$. Using eq.~\eqref{piHat} we have
\begin{equation}
\frac{\mathcal{E}_{(3)}}{\mathcal{E}_{(2)}} \sim \alpha_{\rm H} ^2 \left(\frac{\omega}{H}\right)^3 (h_0^+)^2 \sim \beta^2 \frac{H}{\omega}\;, \hspace{1.5cm} \frac{\mathcal{E}_{(4)}}{\mathcal{E}_{(2)}} \sim \alpha_{\rm H} ^3 \left(\frac{\omega}{H}\right)^6 (h_0^+)^4 \sim \beta^3 h_0^+\;.
\end{equation} 
For $\beta \gtrsim 1$ both these ratios are very small and \eqref{piHat} is valid. (The approximation may not be correct for $\beta \gg 1$, but in this case one can reduce to $\beta \gtrsim 1$ considering weaker---i.e.~farther---GW sources.)  
The background $\hat \pi$ will affect the kinetic term of perturbations $Z_{\mu\nu}$: we have to compare its contribution with the one of GWs, $ \pt_u \Gamma_{\mu\nu} $ (note that for $\tilde{m}_4^2$ the relevant parameter is $\ddot \gamma_{ij}$ rather than $\dot \gamma_{ij}$).
For the cubic self-interaction in \eqref{lagrangian:m4} one gets 
\begin{align}
\frac{\partial^2\hat{\pi}}{\ddot{\gamma}} \sim \alpha_{\rm{H}} \frac{\omega}{H} h_0^+ \sim \beta \frac{H}{\omega} \ll 1 \;.
\end{align}
For the quartic self-interaction in \eqref{lagrangian:m4} one has
\begin{equation}
\frac{(\pt^2 \hat{\pi})^2 \Lambda_\star^3 }{\ddot \gamma \Lambda_c^6} \sim \alpha_{\rm H}^2\left(  \frac{\omega}{H}\right)^4  (h_0^+)^3 \sim \beta^2 h_0^+  \ll 1\;.
\end{equation}
We conclude that one can trust the bound plotted in Fig.~\ref{fig:MBH2}.

As already mentioned, the stability properties of the cubic Galileon interactions (in the absence of GWs) do not hold for the quartic and quintic Galileon \cite{Endlich:2011vg}. This means that these theories are in general unstable in the Vainshtein regime, even before considering GWs. However the two instabilities are quite different: one is only present in the non-linear regime of $\pi$, while the instability we discuss in this paper holds outside the Vainshtein regime and extend to the whole Universe.


\section{Conclusions and prospects}
\label{sec:con} 

We have studied the effect of a large 
GW background on the stability of the Effective Field Theory  of Dark Energy. We have discussed two operators where this effect is relevant:  $m_3^3 (t) \delta g^{00} \delta K$, associated to the dimensionless function $\alpha_{\rm B}$, and   $\frac12 \tilde m_4^2(t) \delta g^{00} \left({}^{(3)}\!R + \delta K_\mu^\nu \delta K_\nu^\mu - \delta K^2  \right)$, associated to $\alpha_{\rm H}$. 
We have first focused the analysis on the former, because this operator remains unconstrained by the perturbative decay of gravitons, since the scale suppressing the coupling $\gamma \pi \pi$ is typically too high \cite{Creminelli:2018xsv}. Moreover, the resonant decay is quenched by the non-linear self-couplings of $\pi$, so that also in this regime there are no conclusive bound on this operator from the decay of GWs \cite{Creminelli:2019nok}. 

The stability of perturbations for this operator is studied in Sec.~\ref{sec:stabArg}. For $c_s < 1 $, perturbations of $\pi$  become necessarily unstable in the presence of a GW background with  
\be 
\beta \sim \frac{| \alpha_{\rm B}  | }{  \alpha c_s^2 } \, \frac{  \omega  }{   H_0 } \, h_0^+  > 1 \;:
\ee
the kinetic matrix $Z^{\mu \nu}$ presents either ghost or gradient instabilities.
These conclusions are not at variance with the well-known  theorem that ensures stability for the DGP model \cite{Nicolis:2004qq} in the absence of GWs. In this case, for $c_s^2 > 1/3$ the theorem can be extended to the $m_3^3$ operator, but
its assumptions break down when a GW background is present. The case $c_s=1$ is also discussed, assuming that the $\pi$ background is linear in the light-cone coordinate $v= t+r$. In this case we find that ghost instabilities are generic. Our conclusions do not extend directly  to the DGP model, because the coupling of the scalar bending mode to tensor modes is different. It would be interesting to verify the stability of the DGP model in the presence of a GW background.

The physical implication of these instabilities is unclear, since the most unstable modes are the closest to the cut-off. Sensible conclusions can only be drawn with the knowledge of the UV completion of the theory. We discussed this in Sec.~\ref{sec:fateofin} by an example unrelated to our theory: a  $U(1)$-symmetric theory for a complex scalar with a mexican-hat potential. In the broken phase, at low energy this system can be described by an effective $P( (\partial \pih)^2 )$-theory for the angular mode $\pih$.  Even if the effective theory presents ghost or gradient instabilities for certain values of $P'$ and $P''$, the UV completion remains perfectly healthy. From this example one can argue that it is not possible to continue the time evolution of the system without knowing the complete theory.

In Sec.~\ref{sec:constraints} we explore for which values of the dimensionless function $\alpha_{\rm B}$ the EFT becomes unstable everywhere in the Universe, losing predictability. The bounds are shown Figs.~\ref{fig:stellar} and \ref{fig:MBH} for different GW sources and roughly correspond to $| \alpha_{\rm B} | \gtrsim  10^{-2}$. They are thus very close to the forecasted limits on this parameter reachable with future large-scale structure observations (see e.g.~\cite{Gleyzes:2015rua,Alonso:2016suf,Noller:2018wyv,Frusciante:2019xia}). For this reason, it would be interesting to improve our analysis considering a more refined estimate of the abundance of massive BH binaries \cite{Bonetti:2017lnj,Bonetti:2018tpf}.  Indeed, the limits obtained from these events are the most interesting, as they correspond to frequencies well below the cut-off of the theory. Of course, a logical possibility is that the EFT breaks down at extremely large scales and it cannot be used to study any GW event. In this scenario also all the classical tests of GR are outside the EFT and one cannot rely on screening mechanisms to explain the success of GR at short scales. 

In Sec.~\ref{sec:m4} we discuss the instability of the operator $\tilde m_4^2$, triggered by a GW background with  
\be 
\beta \sim  \frac{| \alpha_{\rm H}| }{\alpha c_s^2  } \left( \frac{\omega }{H_0}\right)^2 h_0^+ \;.
\ee
This is easier to study because one can neglect non-linearities of $\pi$. The bounds on $\alpha_{\rm H}$ based on $c_T = 1$ are shown in Fig.~\ref{fig:MBH2}: the effective theory becomes unstable for $|\alpha_{\rm H}| \gtrsim 10^{-20}$. These are  much smaller values than those constrained by the perturbative decay. If $c_T = 1$ is relaxed, the combination contrained by our analysis is  $(\tilde{m}_4^2 + m_5^2c_T^2)/\MP^2$, instead of $\tilde{m}_4^2/\MP^2$ (see eq.~(4.15) of \cite{Creminelli:2019nok}).

\renewcommand{\arraystretch}{2}
\begin{table}[t] \centering \normalsize
\begin{adjustbox}{width=1\textwidth}
\begin{tabular}{!{\color{black}\vrule} 
>{\columncolor[HTML]{EFEFEF}}c!{\color{black}\vrule}c!{\color{black}\vrule}c!{\color{black}\vrule}}
  \arrayrulecolor{black}\hline
EFT of DE  operator & $ \frac12 \tilde m_4^2 \, \delta g^{00} \left({}^{(3)}\!R + \delta K_\mu^\nu \delta K_\nu^\mu - \delta K^2  \right)$ & $ m_3^3 \, \delta g^{00} \delta K$   \\
  \arrayrulecolor{black}\hline
  \begin{tabular}{@{}c@{}}  GLPV covariant Lagrangian with $c_T=1$: \\  {\small $ {\mathscr L}= P + Q \Box \phi  + f R -\frac{4 f_{,X}}{X} (\phi^{;\mu} \phi^{;\nu} \phi_{;\mu \nu} \Box \phi  - \phi^{;\mu} \phi_{; \mu \nu} \phi_{; \lambda} \phi^{; \lambda \nu} ) \nonumber $} \end{tabular} &  $ -\frac{2 X f_{,X}}{f}$  & $\frac{2 X f_{,X}}{f} + \frac{\dot \phi X Q_{,X}}{2 H f} $ \\
  \arrayrulecolor{black}\hline
Dimensionless function $ \alpha_i$  & $ \alpha_{\rm H}$ & $ \alpha_{\rm B}$   \\
  \arrayrulecolor{black}\hline
  After conformal transformation &  $\alpha_{\rm H} + 2 \beta_1$ & $\alpha_{\rm B} - \frac{\alpha_{\rm M}}{2} (1-\beta_1) + \beta_1 - \frac{\dot \beta_1}{H} $  \\
  \arrayrulecolor{black}\hline \hline
  Perturbative decay ($\Gamma_{\gamma \rightarrow \pi\pi}/H_0 < 1$) \cite{Creminelli:2018xsv} &  $|\alpha_{\rm H}| \gtrsim 10^{-10}  $ & Irrelevant ($|\alpha_{\rm B}| \gtrsim 10^{10}$)  \\
  \arrayrulecolor{black}\hline
  Narrow resonance ($\beta < 1$, $\beta \omega u >  1$) \cite{Creminelli:2019nok} & \begin{tabular}{@{}c@{}}  $3 \times 10^{-20}  \lesssim  |\alpha_{\rm H} | \lesssim 10^{-17}$ with LIGO-Virgo \\ $ 10^{-16}  \lesssim  |\alpha_{\rm H} | \lesssim 10^{-10}$ with LISA \end{tabular} & \begin{tabular}{@{}c@{}}  Not applicable \\(large non-linearities)\end{tabular}    \\
\arrayrulecolor{black}\hline
 Instability ($\beta>1$, $\beta \omega u >  1$) &  $|\alpha_{\rm H}| \gtrsim 10^{-20}$ (see Fig.~\ref{fig:MBH2}) &  $|\alpha_{\rm B}| \gtrsim 10^{-2}$ (see Fig.~\ref{fig:MBH})  \\
 \arrayrulecolor{black}\hline
\end{tabular}
\end{adjustbox}
\caption{:~Summary of the results of Refs.~\cite{Creminelli:2018xsv,Creminelli:2019nok} and this article (we assume $c_T = 1$).   }
\label{tab1}
\end{table}
In this paper we considered the effects of $\alpha_{\rm B}$ and $\alpha_{\rm H}$ independently but we do not expect that the combination of the two operators can provide better stability properties for $\pi$.
Indeed, the lack of a general theorem for stability in the presence of GWs suggests that our conclusions hold in a more general theory, where both operators are turned on. In particular we expect $\alpha_{\textrm{H}}$ to also contribute to the operator $\dot{\gamma}_{ij}\partial_i\pi\partial_j\pi$: it may be possible then to tune $\alpha_{\textrm{H}}$ and $\alpha_{\textrm{B}}$ to set the operator to zero. However the dominant operator $\ddot{\gamma}_{ij}\partial_i\pi\partial_j\pi$, which has more derivatives, would then lead to instability since it cannot be removed by tuning other parameters.

We summarize these results  and those of Refs.~\cite{Creminelli:2018xsv,Creminelli:2019nok} in Tab.~\ref{tab1}, using different notations. For simplicity, in eqs.~\eqref{starting_action} and \eqref{eq:bh}  we have assumed that our starting theory has a constant effective Planck mass and no higher-derivative operators such as those appearing in DHOST theories \cite{Langlois:2015cwa,Crisostomi:2016czh}. However, our results also apply after a conformal transformation with conformal factor depending on $\phi$ and $X \equiv g^{\mu \nu} \partial_\mu \phi \partial_\nu \phi$: $g_{\mu \nu} \to C(\phi, X) g_{\mu \nu}$. In the fourth line of the table we provide the corresponding parameters to which our analysis applies.

We conclude that for what concerns large-scale structure surveys, the surviving single-field theory that avoids the aforementioned issues is a $k$-essence theory \cite{ArmendarizPicon:2000dh,ArmendarizPicon:2000ah}, modulo the above conformal transformation. In the covariant language, its action  reads
\be
{\mathscr L} = P(\phi, X) + C(\phi, X) R + \frac{6 C_{,X}(\phi, X)^2}{C(\phi, X)}  \phi^{; \mu}   \phi_{; \mu \nu}  \phi_{; \lambda}  \phi^{; \nu \lambda} \;,
\ee 
where the symbol $;$ stands for a covariant derivative.
Note that there is no Vainshtein screening in these theories \cite{Crisostomi:2019yfo}: some other mechanism (see e.g.~\cite{Joyce:2014kja,Brax:2014yla} and references therein) is required to screen the fifth force on astrophysical scales. We leave the study of this regime to future work.

\section*{Acknowledgements}
It is a pleasure to thank E.~Babichev, E.~Barausse, E.~Bellini, S.~Melville, A.~Nicolis, I.~Saltas, I.~Sawicki and A.~Vikman for useful discussions. F.V. acknowledges the Institut Pascal at Universit\'e Paris-Saclay, with the support of the P2I and SPU research departments and  the P2IO Laboratory of Excellence (program Investissements d'avenir ANR-11-IDEX-0003-01 Paris-Saclay and ANR-10-LABX-0038), where this article was finalized.

\appendix 


\section{Deviation from cubic Galileon}
\label{app:cG}
The discussion of Sec.~\ref{sec:stabArg} assumes that the relevant cubic non-linearities are of the form $\tilde{m}_3^3 = - m_3^3$ (as is the case for cubic Galileon interactions). However, one could wonder whether a different choice of operators can make the theory stable around GW backgrounds. To address this possibility, in this appendix we are going to study the stability properties of theories that deviate from the cubic Galileon for the case $c_s<1$.
For concreteness we focus on the case $\tilde m_3^3 = -m_3^3(1+\eta)$, with $\eta \neq 0$ parametrizing such deviations.

The leading non-linear interactions of $\pi$ arising from this coupling are again cubic. The Lagrangian takes then the form
\begin{align}
\mathscr L = -\frac{1}{2}\bar \eta_{\mu \nu} \partial^\mu \pi \partial^\nu \pi -\frac{1}{\Lambda_{\rm B}^3} \square \pi (\partial \pi)^2 + \frac{\eta}{\Lambda_{\rm B}^3}\ddot \pi (\partial_i \pi)^2 + \Gamma_{\mu \nu} \partial^{\mu} \pi \partial^\nu \pi - \frac{\Lambda_{\rm B}^3}{2}\pi\Gamma_{\mu \nu} \Gamma ^{\mu \nu}\;. \label{eq:LagDGPDet2}
\end{align}
Notice that the terms proportional to $\eta$ do not change the couplings with $\Gamma_{\mu\nu}$ and $\Gamma_{\mu\nu}\Gamma^{\mu\nu}$: the operator $(\delta g^{00})^2 \delta K$ yields interactions between $\gamma_{ij}$ and $\pi$ that only start at quartic order.
Straightforwardly, the equation of motion reads
\begin{align}
\bar \square \pi - \frac{2}{\Lambda_{\rm B}^3}\left[ \left( \partial_\mu \partial_\nu \pi\right)^2 - \square \pi ^2 \right] + \frac{2\eta}{\Lambda_{\rm B}^3}\left[ \left(\partial_i\dot{\pi}\right)^2 - \ddot{\pi}\nabla^2\pi \right] -2 \, \Gamma_{\mu \nu} \partial^\mu \partial^\nu \pi - \frac{\Lambda_{\rm B}^3}{2}\Gamma_{\mu \nu} \Gamma ^{\mu \nu} = 0\;.
\end{align}
Following the discussion of Sec.~\ref{sec:csless1}, we have that for $c_s<1$ the solution is a function of $u$ only: $\varphi = \varphi (u)$. In this case one can check that there are no contributions proportional to $\eta$, hence the above equation reduces to \eqref{eq:solXcs}.

At this stage we can compute the kinetic matrix $Z^{\mu\nu}$ for perturbations $\delta \pi$.
By expanding \eqref{eq:LagDGPDet2} at quadratic order we obtain
\begin{align}
Z^{\mu \nu } = - \frac{1}{2}\bar \eta^{\mu \nu} + \Gamma^{\mu \nu} + \frac{2}{\Lambda_{\rm B}^3} \left[\partial^\mu \partial^\nu \hat{\pi}  - \eta^{\mu \nu} \square \hat{\pi} \right] + \eta {\cal R}^{\mu \nu} \;,
\end{align}
where the matrix ${\cal R}^{\mu \nu}$ is defined as 
\begin{equation}
\mathcal R^{\mu \nu} \equiv \frac{1}{\Lambda_{\rm B}^3} \left[
\begin{array}{c|c}
\nabla^2 \hat{\pi} & -\partial_j\dot{\hat{\pi}}   \\ \hline
-\partial_i\dot{\hat{\pi}} &   \ddot{\hat{\pi}} \, \delta_{ij}
\end{array}
\right]\;.
\end{equation}
This expression for $Z^{\mu\nu}$ should be compared with the case $\eta = 0$ of eq.~\eqref{eq:lagpert}.

Using the $u$-dependent solution \eqref{eq:solXcs} for $\varphi(u)$ and the change of variables of eq.~\eqref{eq:derivatives}, one finds the non-vanishing components of $Z^{\mu\nu}$, that are given by
\begin{align}
\label{eq:ZcompCs_eta}
&Z^{00} = \frac{1}{2} + (2 + \eta)  \varphi''(u)  \;,\nonumber \\
&Z^{11} =  - \frac{1}{2} c_s^2+ \Gamma^{11} + \eta \varphi''(u) \;, \nonumber\\
&Z^{22} =  - \frac{1}{2} c_s^2+ \Gamma^{22} + \eta \varphi''(u) \;, \\
&Z^{33} = - \frac{1}{2} c_s^2+ (2 + \eta)  \varphi''(u) \;, \nonumber\\
&Z^{03} = Z^{30} = ( 2 + \eta) \varphi''(u)  \;. \nonumber
\end{align}
Now we can see that with this choice of solution the contributions arising from $\eta$-term are the same in all the entries: $\eta \varphi''$.
To avoid gradient instabilities along $x$, one requires $\eta >0 $ and sufficiently large. However with this choice one clearly encounters ghosts. Hence, for any value of $\eta$ the system remains unstable. 


\section{Interactions in Spatially-Flat Gauge}
\label{app:B}

In this appendix we are going to check that the cubic term $\gamma\gamma\pi$ in the Lagrangian \eqref{InteractionLagrangian}, computed in Newtonian gauge, can be obtained also in spatially-flat gauge up to field redefinitions. The same check for the $\gamma\pi\pi$ interaction was already done in \cite{Creminelli:2019nok}. For simplicity we limit our check to the case in which the contribution of matter is negligible ($\rho_{m} = P_m = 0$); in this case one has $c = - \MP^2 \dot H$ in the action \eqref{starting_action} (see e.g.~\cite{Cheung:2007st,Gubitosi:2012hu}).

Here the metric is decomposed as
\begin{equation}
{\rm d}s^2 = - (1+\delta N)^{2}{\rm d }\tilde t^2 + a(\tilde t)^{2} \left( e^{\gamma} \right)_{ij} ( {\rm d }\tilde x^{i} + \tilde N^{i} {\rm d}\tilde t )( {\rm d }\tilde x^{j} + \tilde N^{j} {\rm d}\tilde t )\;,
\end{equation}
where $\tilde \gamma_{ij}$ is transverse and traceless ($\delta_{ij} \tilde{\gamma}_{ij} = 0$, $\tilde \partial_i \tilde{\gamma}_{ij} = 0$) and the shift vector $\tilde N_{i}$ is decomposed as $\tilde N_{i} = \tilde{\partial}_{i}\psi + \hat{\tilde{N}}_i$ with $\tilde \partial_{i} \hat{\tilde{N}}_{i}=0$ ($\tilde \partial_{\mu} = \partial/\partial \tilde x^{\mu}$). We are going to denote the Goldstone field as $\tilde{\pi}(\tilde{x})$. 

The action in spatially-flat gauge does not contain time derivatives of $\delta N$ and $\tilde{N}_{i}$, which are therefore Lagrange multipliers. Hence, at the perturbative level they are fixed by the constraint equations as (see e.g.~\cite{Creminelli:2019nok,Maldacena:2002vr})
\begin{equation}\label{eq:constraint_deltaN}
\delta N = \frac{m_3^{3}}{m_3^{3} - 2\MP^2 H} \dot{\tilde \pi} + \frac{2 \MP^2 \dot H}{m_3^{3} - 2 \MP^{2} H}\tilde \pi \equiv \alpha_{N} \dot{\tilde{\pi}} + \tilde{\alpha}_{N}\tilde{\pi}
\end{equation}
and
\begin{equation}\label{eq:constraint_psi}
\tilde{\psi} = - \frac{m_3^{3}}{m_3^{3} - 2\MP^2 H} {\tilde \pi} - \frac{3 m_{3}^3 H - 4 H \MP^2 (\dot H \MP^2 -2 m_{2}^4)}{(m_3^3 - 2 \MP^2 H)^2} \frac{a^{2}}{\tilde \nabla^2} \dot{\tilde{\pi}} \equiv \alpha_{\psi} \tilde{\pi} + \tilde{\alpha}_{\psi} \frac{a^{2}}{\tilde{\nabla}^2} \dot{\tilde{\pi}}\;,
\end{equation}
where $\tilde{\nabla}^{2} = \tilde \partial_{i} \tilde\partial_{i}$. Following \cite{Creminelli:2019nok}, the field $\tilde{\pi}$ is canonically normalized as
\begin{equation}\label{eq:tilde_pi_to_canonical}
\tilde \pi =  \frac{2 H \MP ^2 - m_3^3}{\sqrt{2} H \MP [ 3 m_3^6 + 4 \MP^2 (c+ 2 m_2^4) ]^{1/2}}\pi_c\;,
\end{equation}
while the canonical normalization for $\tilde\gamma_{ij}$ is the same as in Newtonian gauge. 

It is straightforward to realize that the vertex $\gamma\gamma\pi$ is not generated by the term $m_3^{3} \delta \tilde g^{00} \delta \tilde K $ in the action (it is not possible to get $\gamma^2$ out of either $\delta g^{00}$ or $\delta K$). 
On the other hand, the sought out vertex is generated by the Einstein-Hilbert term.
In order to simplify the derivation we are going to exploit the fact that, as in Newtonian gauge, tensor perturbations $\tilde \gamma_{ij}$ couple to the metric as a minimally-coupled scalar does (this statement can be verified explicitly by expanding the Einstein-Hilbert term up to cubic order).
Therefore the quadratic Lagrangian for $\tilde \gamma_{ij}$ is 
\begin{align}
\mathscr L &= - \frac{\MP^{2}}{8} \tilde g^{\mu\nu} \tilde \partial_{\mu} \tilde{\gamma}_{ij} \tilde{\partial}_{\nu}\tilde \gamma_{ij}\sqrt{-\tilde g} \nonumber \\
		   & = -a^{3}\frac{\MP^2}{8} \left[ - \dot{\tilde{\gamma}}_{ij}^{2} + \frac{1}{a^{2}} (\tilde{\partial}_{k} \tilde \gamma_{ij})^2 + \delta N \left(  \dot{\tilde{\gamma}}_{ij}^{2} + \frac{1}{a^{2}} (\tilde{\partial}_{k} \tilde \gamma_{ij})^2  \right) +  \frac{2}{a^2} \tilde \partial_k \tilde{\psi} \tilde \partial_k \tilde{\gamma}_{ij} \dot{\tilde{\gamma}}_{ij}  \right]\;.
\end{align}
The first two terms in the last equation are the standard kinetic term for the graviton, and the remaining terms contribute to our cubic vertex. 

Let us focus on these relevant terms. By replacing $\delta N$ and $\tilde \psi$ by the constraints \eqref{eq:constraint_deltaN} and \eqref{eq:constraint_psi} we obtain, after several integrations by part and after dropping terms with less than two derivatives,
\begin{align}\label{eq:Lag_ggp_SF_1}
\mathscr L_{\gamma \gamma \pi} = &-a^{3}\frac{\MP^2}{8} \bigg[ (\alpha_{N} + \alpha_{\psi}) \bigg( \dot{\tilde{\gamma}}_{ij}^2 + \frac{1}{a^2}(\tilde \partial_k \tilde \gamma_{ij})^2 \bigg)\dot{\tilde \pi}  + (\alpha_{\psi} H + \dot{\alpha}_{\psi} + \tilde{\alpha}_{N} +c_s^2 \tilde{\alpha}_{\psi})  \frac{1}{a^2}(\tilde{\partial_k} \tilde \gamma_{ij})^2 \tilde{\pi}   \nonumber \\ 
&  - (3 H \alpha_{\psi } - \dot{\alpha}_{\psi} - \tilde{\alpha}_{N} - c_s^2 \tilde\alpha_{\psi}) \dot{\tilde{\gamma}}_{ij}^2 \,\tilde{\pi}\bigg]\;,
\end{align}
where we have used the linear equations of motion for $\tilde{\gamma}_{ij}$, $ \ddot{\tilde{\gamma}}_{ij} + 3 H \dot{\tilde{\gamma}}_{ij} - \frac{1}{a^{2}}\tilde{\nabla}^2 \tilde{\gamma}_{ij} = 0 $ and for $\tilde{\pi}$, $\ddot{\tilde{\pi}} + 3 H \dot{\tilde{\pi}} - \frac{c_s^2}{a^{2}}\tilde{\nabla}^2 \tilde{\pi} = 0$.
The first term in the above equation vanishes since $\alpha_{N} + \alpha_{\psi} = 0$, as one can see from the definitions of $\alpha_{N}$ and $\alpha_\psi$ in eqs.~\eqref{eq:constraint_deltaN} and \eqref{eq:constraint_psi}. 
 Then, using the expression for $c_s^2$ in eq.~\eqref{Cs}, also the second term of \eqref{eq:Lag_ggp_SF_1} vanishes (notice that our expression for $c_s^2$ assumes $\dot{m}_3 = 0$, but the cancellation works also in the more general case \cite{Gleyzes:2014qga}). Therefore one is left only with the term in the last line, which simplifies to 
\begin{equation}\label{eq:Lag_ggp_SF_2}
\mathscr L_{\gamma\gamma\pi} =  \frac{a^{3}}{2}  \frac{m_3^{3} \MP^2 H}{ 2\MP^2 H - m_3^{3} } \dot{\tilde{\gamma}}_{ij}^2 \, \tilde{\pi}\;.
\end{equation}
After using \eqref{eq:tilde_pi_to_canonical} and \eqref{canonical} to go to canonical normalization for $\tilde \pi$ and $\tilde\gamma_{ij}$, equation \eqref{eq:Lag_ggp_SF_2} matches exactly with the vertex in Newtonian gauge \eqref{InteractionLagrangian}.  


\small{
\bibliography{bib_v3} }
\bibliographystyle{utphys}

\end{document}